\documentclass[12pt]{article}

\usepackage{graphicx}
\usepackage{amsmath}

\def\ltwid{\mathrel{\raise.3ex\hbox{$<$\kern-.75em\lower1ex\hbox{$\sim$}}}}
\def\gtwid{\mathrel{\raise.3ex\hbox{$>$\kern-.75em\lower1ex\hbox{$\sim$}}}}

\def\square{\kern1pt\vbox{\hrule height 1.2pt\hbox{\vrule width 1.2pt\hskip 3pt
   \vbox{\vskip 6pt}\hskip 3pt\vrule width 0.6pt}\hrule height 0.6pt}\kern1pt}
\def\overleftrightarrow#1{\vbox{\ialign{##\crcr
     $\leftrightarrow$\crcr\noalign{\kern-1pt\nointerlineskip}
     $\hfil\displaystyle{#1}\hfil$\crcr}}}

\begin{document}

\begin{titlepage}

\begin{flushright}
ITP-UU-13/19, SPIN-13/13, UFIFT-QG-13-08
\end{flushright}

\vskip 1cm

\begin{center}
{\bf Representing the Graviton Self-Energy on de Sitter Background}
\end{center}

\vskip 1cm

\begin{center}
Katie E. Leonard$^{1*}$, Sohyun Park$^{2\S}$, Tomislav Prokopec$^{3\dagger}$\\
and R. P. Woodard$^{4\ddagger}$
\end{center}

\begin{center}
\it{$^{1}$ Applied Mathematics, Inc., 1622 Route 12, PO Box 637 \\
Gales Ferry, CT 06335, UNITED STATES}
\end{center}

\begin{center}
\it{$^{2}$ Institute for Gravitation \& the Cosmos, Pennsylvania State University \\
University Park, PA 16802, UNITED STATES}
\end{center}

\begin{center}
\it{$^{3}$ Institute of Theoretical Physics, Spinoza Institute and the \\
Center for Extreme Matter and Emergent Phenomena (EMME$\Phi$)\\
Utrecht University, Leuvenlaan 4, 3584 CE Utrecht, THE NETHERLANDS}
\end{center}

\begin{center}
\it{$^{4}$ Department of Physics, University of Florida \\
Gainesville, FL 32611, UNITED STATES}
\end{center}

\vskip .5cm

\begin{center}
ABSTRACT
\end{center}

We derive a noncovariant but simple representation for the
self-energy of a conformally transformed graviton field on the
cosmological patch of de Sitter. Our representation involves four
structure functions, as opposed to the two that would be necessary
for a manifestly de Sitter invariant representation. We work out
what the four structure functions are for the one loop correction
due to a massless, minimally coupled scalar. And we employ the
result to work out what happens to dynamical gravitons.

\begin{flushleft}
\end{flushleft}

\vskip 1cm

\begin{flushleft}
$^*$ e-mail: keleonard262@gmail.com \\
$^{\S}$ e-mail: spark@gravity.psu.edu \\
$^{\dagger}$ e-mail: T.Prokopec@uu.nl \\
$^{\ddagger}$ e-mail: woodard@phys.ufl.edu
\end{flushleft}

\end{titlepage}

\section{Introduction}\label{intro}

Inflation produces vast ensembles of infrared scalars and gravitons
which are thought to be the source of primordial perturbations
\cite{perts}. The primary perturbations are a tree order effect,
which means that how they interact among themselves and with other
particles is a loop correction. One studies these loop effects by
first computing the appropriate 1PI (one-particle-irreducible)
2-point function and then using it to quantum-correct the linearized
effective field equation for the particle in question
\cite{phi4,SQED,PP,Yukawa,MW1,SPM,KW1,PW1,PW2,LW,PMTW}.

If the particle we are studying has nonzero spin then its 1PI
2-point function must carry tensor or spinor indices. For example,
the self-energy of a Dirac fermion has 16 bi-spinor components, the
vacuum polarization possesses 16 bi-vector components, while the
graviton self-energy contains 100 bi-tensor components. Although it
would not be wrong to report results for each component separately,
experience with quantum field theory on flat space shows that this
is wasteful and that it obscures important features of the dynamics.
For example, the combination of gauge invariance and Poincar\'e
invariance implies that the vacuum polarization of flat space can be
expressed in terms of a single scalar structure function,
\begin{equation}
i \Bigl[\mbox{}^{\mu} \Pi^{\nu}_{\rm flat} \Bigr](x;x') = \Bigl[
\partial^{\mu} \partial^{\nu} \!-\! \eta^{\mu\nu} \partial^2 \Bigr]
i \Pi\Bigl( (x \!-\! x')^2 \Bigr) \; , \label{flatPi}
\end{equation}
where $\eta^{\mu\nu}$ is the Minkowski metric and $(x - x')^2 \equiv
\eta_{\mu\nu} (x - x')^{\mu} (x - x')^{\nu}$.

How much the 1PI 2-point function can be simplified depends partly
upon linearized gauge invariances and partly on isometries. The
background geometry appropriate for inflationary cosmology is
homogeneous, isotropic and spatially flat,
\begin{equation}
ds^2 = -dt^2 + a^2(t) d\vec{x} \!\cdot\! d\vec{x} = a^2 \Bigl[
-d\eta^2 + d\vec{x} \!\cdot\! d\vec{x} \Bigr] \; . \label{FRW}
\end{equation}
For many purposes it is also desirable to take the de Sitter limit
in which the scale factor becomes $a = -1/H\eta$, with constant $H$.
That would introduce four additional isometries, however, one has to
bear in mind that de Sitter can only be an approximation when
studying primordial inflation. There can also be important de Sitter
breaking effects from inflationary scalars
\cite{AF,phi4,SQED,Yukawa} and gravitons
\cite{TW1,RPW,MTW,KMW,MMTW,MW1,LW,PMTW}. And even when exact de
Sitter invariance is present the cost of making it manifest can be
prohibitive \cite{LPW1,PW1}.

Our goal is to develop a simple form for representing the graviton
self-energy on de Sitter background. The contributions to this from
many sorts of matter fields are de Sitter invariant, and a fully de
Sitter invariant representation using two structure functions was
derived in a previous study of the one loop graviton self-energy
from a massless, minimally coupled scalar \cite{PW1}. However, this
representation turned out to be horrifically complicated \cite{PW1},
and tedious to employ \cite{PW2}.

We also suspect the de Sitter invariant representation for gravitons
might give misleading results. It is disturbingly similar to what we
recently found for the de Sitter invariant contribution to the
vacuum polarization from a massive scalar \cite{LPW1}. The original
computation of this effect was reported using a noncovariant
representation in which there are two structure functions and only
the isometries of homogeneity and isotropy are manifest \cite{PP}.
With that representation the dynamics are transparent and it was
simple to show that dynamical photons become massive. With our de
Sitter invariant representation the dynamics are cumbersome and the
mass contribution to the effective field equation takes the form of
an integral over the initial value surface. These surface terms are
usually irrelevant but we were (at length) able to recognize this
one as a local Proca mass term using Green's second identity
\cite{LPW1}. Had that surface term been discarded we would have
erroneously reached the null conclusion which {\it was} reached for
the graviton using the same, cumbersome and confusing de Sitter
invariant formalism and dropping the same sort of surface term
\cite{PW2}.

In section~\ref{conformal} we show that linearized gauge invariance,
with only the isometries of homogeneity and isotropy, results in
four structure functions for the graviton self-energy, rather than
the two of a fully de Sitter invariant representation. We explicitly
construct the 4-function representation in section~\ref{project}. In
section~\ref{structure} we work out the four structure functions for
the one loop contribution from a massless, minimally coupled scalar.
The new representation is employed in section~\ref{gravitons} to
re-examine the question of quantum corrections to dynamical
gravitons. Our conclusions comprise section~\ref{discuss}.

\section{Counting and Conformal Rescaling}\label{conformal}

The point of this section is to count the number and type of
structure functions which are required to represent the graviton
self-energy when we relax the assumption of full de Sitter
invariance to just the cosmological isometries of homogeneity and
isotropy. We begin by describing the coordinate system of the
cosmological patch and the natural basis vectors on it. The number
of structure functions is just the number of homogeneous and
isotropic basis tensors minus the number of constraints implied by
transversality. An important subtlety is that the simple
representation we aim to derive is for the conformally rescaled
graviton self-energy. The section closes by laying out precisely
what transversality implies for this quantity.

\subsection{Coordinates and basis vectors}

We work on open conformal coordinates (in $D$ spacetime dimensions
to facilitate dimensional regularization),
\begin{equation}
ds^2 = a^2(\eta) \Bigl[ -d\eta^2 + d\vec{x} \!\cdot\! d\vec{x}
\Bigr] \qquad , \qquad a(\eta) = -\frac1{H \eta} \; ,
\end{equation}
where the coordinate ranges are,
\begin{equation}
-\infty < x^0 \equiv \eta < 0 \qquad , \qquad -\infty < x^i <
+\infty \quad, \quad i = 1, 2, \dots , D\!-\!1 \; .
\end{equation}
When a bi-tensor density such as the graviton self-energy is de
Sitter invariant it can be expressed using the invariant length
$\ell(x;x')$. For quantum field theory computations it is most
convenient to employ the de Sitter length function $y(x;x') \equiv 4
\sin^2[\frac12 H \ell(x;x')]$,
\begin{equation}
y(x;x') \equiv H^2 a a' \Bigl[ \Vert \vec{x} \!-\! \vec{x}' \Vert^2
- (\vert \eta \!-\! \eta'\vert - i \varepsilon )^2 \Bigr] \; ,
\label{ydef}
\end{equation}
where $a \equiv a(\eta)$ and $a' \equiv a(\eta')$. When only
homogeneity and isotropy are present one must allow additional
dependence upon two combinations of the scale factors,
\begin{equation}
u(x;x') \equiv \ln(a a') \qquad , \qquad v(x;x') \equiv \ln\Bigl(
\frac{a}{a'} \Bigr) \; . \label{uandv}
\end{equation}

A convenient basis of de Sitter invariant bi-tensors can be formed
using products of the metrics at $x^{\mu}$ and ${x'}^{\mu}$, along
with the first three derivatives of $y(x;x')$,
\begin{eqnarray}
\partial_{\mu} y = a H \Bigl( \delta^0_{~\mu} y \!+\! 2 a' H
\Delta x_{\mu} \Bigr) \quad , \quad
\partial'_{\nu} y = a' H \Bigl( \delta^0_{~\nu} y \!-\! 2 a H \Delta
x_{\nu} \Bigr) \; , \qquad \label{basis1} \\
\partial_{\mu} \partial'_{\nu} y = a a' H^2 \Bigl( \delta^0_{~\mu}
\delta^0_{~\nu} y \!-\! 2 a \delta^0_{~\mu} H \Delta x_{\nu} \!+\! 2
a' \delta^0_{~\nu} H \Delta x_{\mu} \!-\! 2 \eta_{\mu\nu} \Bigr) \;
, \qquad \label{basis2}
\end{eqnarray}
where $\Delta x_{\mu} \equiv \eta_{\mu\nu} (x - x')^{\nu}$. It is
straightforward to show that covariant derivatives and/or
contractions of these basis tensors produce only tensors within the
basis \cite{KW2}. When only homogeneity and isotropy are present one
must include the first derivatives of $u(x;x')$ \cite{MTW},
\begin{equation}
\partial_{\mu} u = a H \delta^0_{~\mu} \qquad , \qquad
\partial'_{\nu} u = a' H \delta^0_{~\nu} \; . \label{basis3}
\end{equation}
Derivatives of $v(x;x')$ are unnecessary because $\partial_{\mu} v =
+\partial_{\mu} u$ and $\partial'_{\nu} v = -\partial'_{\nu} u$.
Acting covariant derivatives on any element of
(\ref{basis1}-\ref{basis3}), or contracting any two elements,
produces sums of products of more basis elements \cite{MTW}.

\subsection{Counting the structure functions}

If the full metric is $g^{\rm full}_{\mu\nu}$ and the metric of the
de Sitter background is $g_{\mu\nu} = a^2 \eta_{\mu\nu}$ then the
graviton field of the invariant representation is,
\begin{equation}
\chi_{\mu\nu}(x) \equiv \frac{ g^{\rm full}_{\mu\nu}(x) \!-\!
g_{\mu\nu}(x)}{\kappa} \qquad , \qquad \kappa^2 \equiv 16 \pi G \; .
\label{mathgrav}
\end{equation}
The self-energy of this field is a transverse bi-tensor density,
\begin{equation}
D_{\mu} \Bigl[\mbox{}^{\mu\nu}
\Sigma^{\rho\sigma}_{\chi}\Bigr](x;x') = 0 = D'_{\rho}
\Bigl[\mbox{}^{\mu\nu} \Sigma^{\rho\sigma}_{\chi}\Bigr](x;x') \; ,
\label{transverse}
\end{equation}
where $D_{\mu}$ and $D'_{\rho}$ stand for the covariant derivatives
with respect to $x^{\mu}$ and ${x'}^{\rho}$ computed using the
affine connection of the de Sitter background,
\begin{equation}
\Gamma^{\rho}_{~ \mu\nu}(x) = a H \Bigl( \delta^{\rho}_{~ \mu}
\delta^0_{~ \nu} \!+\! \delta^{\rho}_{~ \nu} \delta^0_{~ \mu} \!-\!
\eta^{\rho 0} \eta_{\mu\nu} \Bigr) = \delta^{\rho}_{~\mu} u_{,\nu}
\!+\! \delta^{\rho}_{~\nu} u_{,\mu} - u^{,\rho} g_{\mu\nu} \; .
\label{Gamma}
\end{equation}
The self-energy is also invariant under interchange of coordinates
and index groups,
\begin{equation}
-i\Bigl[ \mbox{}^{\mu\nu} \Sigma^{\rho\sigma}_{\chi}\Bigr](x;x') =
-i \Bigl[ \mbox{}^{\rho\sigma} \Sigma^{\mu\nu}_{\chi}\Bigr](x';x) \;
. \label{reflection}
\end{equation}

If the graviton self-energy is manifestly de Sitter invariant it
must consist of a linear combination of five tensors,
\begin{eqnarray}
D^{\mu} {D'}^{(\rho} y {D'}^{\sigma )} D^{\nu} y \;\; , \;\;
D^{(\mu} y D^{\nu)} {D'}^{(\rho} y {D'}^{\sigma)} y \;\; , \;\;
D^{\mu} y D^{\nu} y {D'}^{\rho} y {D'}^{\sigma} y \; , \nonumber \\
H^2 \Bigl(g^{\mu\nu} {D'}^{\rho} y {D'}^{\sigma} y \!+\! D^{\mu} y
D^{\nu} y \, {g'}^{\rho\sigma}\Bigr) \;\; , \;\; H^4 g^{\mu\nu}
{g'}^{\rho\sigma} \; . \label{inv5}
\end{eqnarray}
Here and henceforth indices which are enclosed in parentheses are
symmetrized. Transversality (\ref{transverse}) means the covariant
divergence $D_{\mu}$ vanishes, which implies relations proportional
to the three tensors,
\begin{equation}
D^{\nu} {D'}^{(\rho} y {D'}^{\sigma)} y \;\; , \;\; D^{\nu} y
{D'}^{\rho} y {D'}^{\sigma} y \;\; , \;\; D^{\nu} y \,
{g'}^{\rho\sigma} \; . \label{invdiv}
\end{equation}
Hence we require $5 - 3 = 2$ structure functions to make de Sitter
invariance manifest when it is present. One of these is associated
with a transverse-traceless tensor structure that mixes the
$x^{\mu}$ and ${x'}^{\mu}$ index groups while the other tensor
structure is diagonal \cite{PW1}.

If the only isometries are homogeneity and isotropy the most general
reflection invariant bi-tensor requires nine more tensors in
addition to those of (\ref{inv5}) \cite{KMW},
\begin{eqnarray}
\lefteqn{\Bigl( D^{(\mu} y D^{\nu)} {D'}^{(\rho} y {D'}^{\sigma)} u
\!+\! D^{(\mu} u D^{\nu)} {D'}^{(\rho} y {D'}^{\sigma)} y \Bigr) \;\; ,
\;\; D^{(\mu} u D^{\nu)} {D'}^{(\rho} y {D'}^{\sigma)} u \; , } \nonumber \\
& & \hspace{-.5cm} \Bigl( D^{\mu} y D^{\nu} y {D'}^{\rho} u
{D'}^{\sigma} u \!+\! D^{\mu} u D^{\nu} u {D'}^{\rho} y {D'}^{\sigma} y
\Bigr) \;\; , \;\; D^{(\mu} y D^{\nu)} u {D'}^{(\rho} y {D'}^{\sigma)} u \; , \nonumber \\
& & \hspace{-.5cm} \Bigl( D^{(\mu} y D^{\nu)} u {D'}^{\rho} u
{D'}^{\sigma} u \!+\! D^{\mu} u D^{\nu} u {D'}^{(\rho} u {D'}^{\sigma)} y
\Bigr) \;\; , \;\; D^{\mu} u D^{\nu} u {D'}^{\rho} u {D'}^{\sigma} u \; , \nonumber \\
& & \hspace{-.5cm} H^2 \Bigl( D^{(\mu} y D^{\nu)} u
{g'}^{\rho\sigma} \!+\! g^{\mu\nu} {D'}^{(\rho} y {D'}^{\sigma)} u
\Bigr) \;\; , \;\; H^2 \Bigl( D^{\mu} u D^{\nu} u {g'}^{\rho\sigma}
\!+\! g^{\mu\nu} {D'}^{\rho} u {D'}^{\sigma} u \Bigr) \; , \nonumber \\
& & H^2 \Bigl( D^{\mu} u D^{\nu} u {g'}^{\rho\sigma} \!-\!
g^{\mu\nu} {D'}^{\rho} u {D'}^{\sigma} u \Bigr) \; . \qquad
\label{noninv9}
\end{eqnarray}
Transversality (\ref{transverse}) implies relations proportional to
seven tensors in addition to those of (\ref{invdiv}),
\begin{eqnarray}
\lefteqn{ D^{\nu} {D'}^{(\rho} y {D'}^{\sigma)} u \;\; , \;\;
D^{\nu} y {D'}^{(\rho} y {D'}^{\sigma)} u \;\; , \;\; D^{\nu} y
{D'}^{\rho} u {D'}^{\sigma} u \; , } \nonumber \\
& & \hspace{-.5cm} D^{\nu} u {D'}^{\rho} y {D'}^{\sigma} y \;\; ,
\;\; D^{\nu} u {D'}^{(\rho} y {D'}^{\sigma)} u \;\; , \;\; D^{\nu} u
{D'}^{\rho} u {D'}^{\sigma} u \;\; , \;\; D^{\nu} u \,
{g'}^{\rho\sigma} \; . \qquad \label{nondiv}
\end{eqnarray}
Hence there must be $14 - 10 = 4$ structure functions when only
homogeneity and isotropy are manifest. We might guess that two of
them will be associated with transverse-traceless tensor structures
which mix index groups while the remaining two are diagonal.

\subsection{Conformal rescaling}

Although mathematical physicists prefer to consider the ``graviton
field'' to be the quantity $\chi_{\mu\nu}$ defined in expression
(\ref{mathgrav}), it has long been known that the simplest Feynman
rules arise for the conformally rescaled graviton field \cite{TW1},
\begin{equation}
h_{\mu\nu}(x) \equiv \frac{g^{\rm full}_{\mu\nu}(x) \!-\!
g_{\mu\nu}(x)}{\kappa a^2} = a^{-2} \times \chi_{\mu\nu}(x) \; .
\label{htochi}
\end{equation}
This is the variable for which all of the existing fully
dimensionally regulated graviton loop computations have been made
\cite{MW1,SPM,KW1,LW,PMTW}. Expression (\ref{htochi}) fixes the
relation between the propagator of $\chi_{\mu\nu}$ and that of
$h_{\mu\nu}$,
\begin{eqnarray}
\lefteqn{i \Bigl[ \mbox{}_{\mu\nu}
\Delta^{\chi}_{\rho\sigma}\Bigr](x;x') \equiv \Bigl\langle \Omega_0
\Bigl\vert T\Bigl[ \chi_{\mu\nu}(x)
\chi_{\rho\sigma}(x') \Bigr] \Bigr\vert \Omega_0 \Bigr\rangle \; , } \\
& & = (a a')^2 \times \Bigl\langle \Omega_0 \Bigl\vert T\Bigl[
h_{\mu\nu}(x) h_{\rho\sigma}(x') \Bigr] \Bigr\vert \Omega_0
\Bigr\rangle \equiv (a a')^2 \times i \Bigl[ \mbox{}_{\mu\nu}
\Delta_{\rho\sigma}\Bigr](x;x') \; . \qquad \label{props}
\end{eqnarray}

To infer the corresponding relation between the self-energies of
$\chi_{\mu\nu}$ and $h_{\mu\nu}$, it suffices to compare the one
loop corrections to the full propagators. For the field
$\chi_{\mu\nu}$ we have,
\begin{eqnarray}
\lefteqn{i\Bigl[\mbox{}_{\mu\nu}
\Delta^{\chi1}_{\rho\sigma}\Bigr](x;x') } \nonumber \\
& & \hspace{-.3cm} = \int \!\! d^Dz \! \int \!\! d^Dz' \, i\Bigl[
\mbox{}_{\mu\nu} \Delta^{\chi}_{\alpha\beta}\Bigr](x;z) \times -i
\Bigl[\mbox{}^{\alpha\beta}
\Sigma^{\kappa\lambda}_{\chi}\Bigr](z;z') \times
i\Bigl[\mbox{}_{\kappa\lambda}
\Delta^{\chi}_{\rho\sigma}\Bigr](z';x') \; . \qquad \label{chi1}
\end{eqnarray}
The corresponding expression for the one loop correction to the
$h_{\mu\nu}$ propagator is,
\begin{eqnarray}
\lefteqn{i\Bigl[\mbox{}_{\mu\nu}
\Delta^1_{\rho\sigma}\Bigr](x;x') } \nonumber \\
& & \hspace{-.3cm} = \int \!\! d^Dz \! \int \!\! d^Dz' \, i\Bigl[
\mbox{}_{\mu\nu} \Delta_{\alpha\beta}\Bigr](x;z) \times -i
\Bigl[\mbox{}^{\alpha\beta} \Sigma^{\kappa\lambda}\Bigr](z;z')
\times i\Bigl[\mbox{}_{\kappa\lambda}
\Delta_{\rho\sigma}\Bigr](z';x') \; . \qquad \label{h1}
\end{eqnarray}
Because expression (\ref{chi1}) must be $(a a')^2$ times expression
(\ref{h1}) the two self-energies must be related as,
\begin{equation}
-i \Bigl[ \mbox{}^{\mu\nu} \Sigma_{\chi}^{\rho\sigma}\Bigr](x;x') =
(a a')^{-2} \times -i \Bigl[ \mbox{}^{\mu\nu}
\Sigma^{\rho\sigma}\Bigr](x;x') \; . \label{sigma}
\end{equation}

\subsection{Consequences of transversality}

The self-energy of $\chi_{\mu\nu}$ is a bi-tensor density. This
means it can be expressed as proper bi-tensor times the measure
factors from $x^{\mu}$ and ${x'}^{\mu}$,
\begin{equation}
-i \Bigl[ \mbox{}^{\mu\nu} \Sigma_{\chi}^{\rho\sigma}\Bigr](x;x') =
\sqrt{-g(x)} \, \sqrt{-g(x')} \times \Bigl[\mbox{}^{\mu\nu}
T^{\rho\sigma}\Bigr](x;x') \; . \label{truetens}
\end{equation}
Transversality (\ref{transverse}) means that the covariant
divergence of $[\mbox{}^{\mu\nu} T^{\rho\sigma}](x;x')$ vanishes,
\begin{equation}
D_{\mu} \Bigl[\mbox{}^{\mu\nu} T^{\rho\sigma}\Bigr] = \partial_{\mu}
\Bigl[\mbox{}^{\mu\nu} T^{\rho\sigma} \Bigr] +
\Gamma^{\alpha}_{~\alpha \mu} \Bigl[\mbox{}^{\mu\nu} T^{\rho\sigma}
\Bigr] + \Gamma^{\nu}_{~\mu\alpha} \Bigl[ \mbox{}^{\mu\alpha}
T^{\rho\sigma}\Bigr] = 0 \; .
\end{equation}
Because $\partial_{\mu} \sqrt{-g} = \sqrt{-g} \,
\Gamma^{\alpha}_{~\alpha \mu}$ we can re-express transversality
(\ref{transverse}) using the ordinary derivative,
\begin{equation}
\partial_{\mu} \Bigl[\mbox{}^{\mu\nu}
\Sigma_{\chi}^{\rho\sigma}\Bigr](x;x') +
\Gamma^{\nu}_{~\alpha\beta}(x) \Bigl[\mbox{}^{\alpha\beta}
\Sigma_{\chi}^{\rho\sigma}\Bigr](x;x') = 0 \; .
\end{equation}
Now make use of relations (\ref{Gamma}) and (\ref{sigma}) to
conclude that the self-energy of the field $h_{\mu\nu}$ obeys,
\begin{equation}
\partial_{\mu} \Bigl[\mbox{}^{\mu\nu}
\Sigma^{\rho\sigma}\Bigr](x;x') - D^{\nu} u \, g_{\alpha\beta}(x)
\times \Bigl[\mbox{}^{\alpha\beta} \Sigma^{\rho\sigma}\Bigr](x;x') =
0 \; . \label{ourtrans}
\end{equation}

\section{The Four Projectors}\label{project}

The aim of this section is to derive an explicit expression for the
self-energy of $h_{\mu\nu}$ in terms of the four structure functions
we have just seen are needed when only homogeneity and isotropy are
present. We are guided by two facts, the first of which is the form
taken by the flat space limit,
\begin{equation}
-i \Bigl[\mbox{}^{\mu\nu} \Sigma^{\rho\sigma}_{\rm flat} \Bigr] =
\Pi^{\mu\nu} \Pi^{\rho\sigma} f_0\Bigl( (x \!-\! x')^2 \Bigr) +
\Bigl[ \Pi^{\mu (\rho} \Pi^{\sigma ) \nu} \!-\! \frac{\Pi^{\mu\nu}
\Pi^{\rho\sigma}}{D \!-\! 1} \Bigr] f_2\Bigl( (x \!-\! x')^2 \Bigr)
\; , \label{flatsig}
\end{equation}
where $\Pi^{\mu\nu} \equiv \partial^{\mu} \partial^{\nu} -
\eta^{\mu\nu} \partial^2$. The structure functions $f_0$ and $f_2$
are usually labeled {\it spin zero} and {\it spin two},
respectively. The second fact is the simple noncovariant form of the
vacuum polarization on de Sitter \cite{SQED,PP,LPW1},
\begin{equation}
i\Bigl[ \mbox{}^{\mu} \Pi^{\nu}\Bigr] = \Bigl( \eta^{\mu\nu}
\eta^{\rho\sigma} \!-\! \eta^{\mu\sigma} \eta^{\nu\rho}\Bigr) \,
\partial_{\rho} \partial'_{\sigma} F(x;x') + \Bigl(
\overline{\eta}^{\mu\nu} \overline{\eta}^{\rho\sigma} \!-\!
\overline{\eta}^{\mu\sigma} \overline{\eta}^{\nu\sigma} \Bigr) \,
\partial_{\rho} \partial'_{\sigma} G(x;x') \; , \label{vacpol}
\end{equation}
where an overlined tensor indicates suppression of its temporal
components, $\overline{\eta}^{\mu\nu} \equiv \eta^{\mu\nu} +
\delta^0_{~\mu} \delta^0_{~\nu}$. The structure function $F(x;x')$
is the one which survives in the flat space limit; $G(x;x')$ is less
divergent and vanishes with $H$.

These two hints motivate a representation of the form,
\begin{eqnarray}
\lefteqn{-i\Bigl[ \mbox{}^{\mu\nu} \Sigma^{\rho\sigma}\Bigr](x;x') =
\mathcal{F}^{\mu\nu}(x) \times \mathcal{F}^{\rho\sigma}(x') \Bigl[
F_0(x;x') \Bigr] } \nonumber \\
& & \hspace{-.5cm} + \mathcal{G}^{\mu\nu}(x) \times
\mathcal{G}^{\rho\sigma}(x') \Bigl[ G_0(x;x') \Bigr] +
\mathcal{F}^{\mu\nu\rho\sigma} \Bigl[ F_2(x;x') \Bigr] +
\mathcal{G}^{\mu\nu\rho\sigma} \Bigl[ G_2(x;x') \Bigr] \; . \qquad
\label{ourform}
\end{eqnarray}
The idea is that the two ``F'' terms should represent appropriate de
Sitter generalizations of the flat space result (\ref{flatsig})
while the two ``G'' terms should be essentially ``spatial'' like the
$G$ term in the vacuum polarization (\ref{vacpol}). Each of the four
terms should separately respect transversality (\ref{ourtrans}). So
the second order differential operators $\mathcal{F}^{\mu\nu}$ and
$\mathcal{G}^{\mu\nu}$  must obey,
\begin{equation}
\partial_{\mu} \mathcal{F}^{\mu\nu} + a H \delta^{\nu}_{~ 0}
\eta_{\alpha\beta} \mathcal{F}^{\alpha\beta} = 0 =
\partial_{\mu} \mathcal{G}^{\mu\nu} + a H \delta^{\nu}_{~ 0}
\eta_{\alpha\beta} \mathcal{G}^{\alpha\beta} \; . \label{scalar}
\end{equation}
The differential operators $\mathcal{F}^{\mu\nu\rho\sigma}$ and
$\mathcal{G}^{\mu\nu\rho\sigma}$ should each contain two primed and
two unprimed derivatives and should each be separately transverse
and traceless,
\begin{equation}
\partial_{\mu} \mathcal{F}^{\mu\nu\rho\sigma} = 0 = \partial_{\mu}
\mathcal{G}^{\mu\nu\rho\sigma} \qquad , \qquad \eta_{\mu\nu}
\mathcal{F}^{\mu\nu\rho\sigma} = 0 = \eta_{\mu\nu}
\mathcal{G}^{\mu\nu\rho\sigma} \; . \label{tensor}
\end{equation}

\subsection{The scalar projectors $\mathcal{F}^{\mu\nu}$ and
$\mathcal{G}^{\mu\nu}$}

We construct the two scalar projectors by making an initial ansatz
and then using the transversality relation (\ref{scalar}) to
determine the free parameters. Because we want the flat space limit
of $\mathcal{F}^{\mu\nu}$ to be $\Pi^{\mu\nu} = \partial^{\mu}
\partial^{\nu} - \eta^{\mu\nu} \partial^2$, our ansatz for it is,
\begin{equation}
\mathcal{F}^{\mu\nu} = \partial^{\mu} \partial^{\nu} \!+\! 2 f_1 a H
\delta^{(\mu}_{~0} \partial^{\nu)} \!+\! f_2 a^2 H^2 \delta^{\mu}_{~
0} \delta^{\nu}_{~0} - \eta^{\mu\nu} \Bigl[ \partial^2 \!+\! f_3 a H
\partial_0 \!+\! f_4 a^2 H^2\Bigr] \; . \label{Fansatz}
\end{equation}
Equation (\ref{scalar}) determines $f_1 = f_3 = f_4 = (D-1)$ and
$f_2 = (D-2)(D-1)$. Because the structure functions tend to carry a
factor of $a^{D-2}$ it is useful to note,
\begin{equation}
\mathcal{F}^{\mu\nu} = a^{D-2} \Biggl[ \partial^{\mu} \partial^{\nu}
\!+\! 2 a H \delta^{(\mu}_{~0} \partial^{\nu)} \!-\! \eta^{\mu\nu}
\Bigl[ \partial^2 \!-\! (D \!-\! 3) a H \partial_0 \!+\! (D \!-\! 1)
a^2 H^2 \Bigr] \Biggr] a^{-(D-2)} \! . \label{rescale}
\end{equation}
In this form we can express the trace in terms of the invariant
scalar d'Alem\-bert\-ian $\square = a^{-D} \partial_{\mu} (a^{D-2}
\eta^{\mu\nu} \partial_{\nu})$,
\begin{equation}
a^D \times \mathcal{F} \times a^{-(D-2)} \equiv \eta_{\mu\nu}
\mathcal{F}^{\mu\nu} = -(D \!-\!1) a^D \times \Bigl[ \square \!+\! D
H^2 \Bigr] \times a^{-(D-2)} \; . \label{Ftrace}
\end{equation}

Because we want $\mathcal{G}^{\mu\nu}$ to be ``essentially spatial''
our ansatz for it is,
\begin{equation}
\mathcal{G}^{\mu\nu} = \overline{\partial}^{\mu}
\overline{\partial}^{\nu} \!+\! 2 g_1 a H \delta^{(\mu}_{~0}
\overline{\partial}^{\nu)} \!+\! g_2 a^2 H^2 \delta^{\mu}_{~ 0}
\delta^{\nu}_{~0} - \overline{\eta}^{\mu\nu} \Bigl[ \nabla^2 \!+\!
g_3 a H \partial_0 \!+\! g_4 a^2 H^2\Bigr] \; , \label{Gansatz}
\end{equation}
where we remind the reader that a line over a vector indicates
suppression of its temporal components, $\overline{\partial}^{\mu}
\equiv \partial^{\mu} + \delta^{\mu}_{~0} \partial_0$. Equation
(\ref{scalar}) determines $g_1 = g_3 = g_4 = (D-2)$ and $g_2 =
(D-2)(D-1)$. The trace of $\mathcal{G}^{\mu\nu}$ is,
\begin{eqnarray}
\lefteqn{a^D \times \mathcal{G} \times a^{-(D-2)} \equiv
\eta_{\mu\nu} \mathcal{G}^{\mu\nu} \; , } \\
& & \hspace{.5cm} = -(D \!-\! 2) a^{D} \Bigl[ \frac{\nabla^2}{a^2}
\!+\! (D \!-\! 1) \frac{H}{a} \, \partial_0 \!+\! D(D\!-\! 1) H^2 \Bigr]
\times a^{-(D-2)} \; . \label{Gtrace} \qquad
\end{eqnarray}

\subsection{The tensor projectors $\mathcal{F}^{\mu\nu\rho\sigma}$
and $\mathcal{G}^{\mu\nu\rho\sigma}$}

Our technique for constructing transverse-traceless projectors is a
variation of the one employed in the de Sitter invariant
construction \cite{PW1}. We begin by expanding the Weyl tensor of
the conformally transformed metric,
\begin{equation}
\widetilde{g}_{\mu\nu} \equiv \eta_{\mu\nu} + \kappa h_{\mu\nu}
\qquad \Longrightarrow \qquad
\widetilde{C}_{\alpha\beta\gamma\delta} \equiv
\mathcal{C}_{\alpha\beta\gamma\delta}^{~~~~~ \mu\nu} \times \kappa
h_{\mu\nu} + O(\kappa^2 h^2) \; .
\end{equation}
The second order differential operator
$\mathcal{C}_{\alpha\beta\gamma\delta}^{~~~~~ \mu\nu}$ is,
\begin{eqnarray}
\lefteqn{ \mathcal{C}_{\alpha\beta\gamma\delta}^{~~~~~ \mu\nu} =
\mathcal{D}_{\alpha\beta\gamma\delta}^{~~~~~ \mu\nu} - \frac1{D
\!-\! 2} \Bigl[ \eta_{\alpha\gamma} \mathcal{D}_{\beta\delta}^{~~~
\mu\nu} \!-\! \eta_{\gamma\beta} \mathcal{D}_{\delta\alpha}^{~~~
\mu\nu} } \nonumber \\
& & \hspace{3cm} + \eta_{\beta\delta}
\mathcal{D}_{\alpha\gamma}^{~~~ \mu\nu} \!-\! \eta_{\delta\alpha}
\mathcal{D}_{\gamma\beta}^{~~~ \mu\nu} \Bigr] +
\frac{(\eta_{\alpha\gamma} \eta_{\beta\delta} \!-\!
\eta_{\alpha\delta} \eta_{\beta\gamma}) \mathcal{D}^{\mu\nu}}{(D
\!-\! 1)(D \!-\! 2)} \; , \qquad \label{calC}
\end{eqnarray}
where we can express $\mathcal{D}_{\alpha\beta\gamma\delta}^{~~~~~
\mu\nu}$ in terms of the Minkowski metric and the partial derivative
operator,
\begin{eqnarray}
\mathcal{D}_{\alpha\beta\gamma\delta}^{~~~~~ \mu\nu} & \equiv &
-\frac12 \Bigl( \delta^{(\mu}_{~~ \alpha} \delta^{\nu )}_{~ \gamma}
\partial_{\beta} \partial_{\delta} \!-\! \delta^{(\mu}_{~~ \gamma}
\delta^{\nu )}_{~\beta} \partial_{\delta} \partial_{\alpha} \!+\!
\delta^{(\mu}_{~~ \beta} \delta^{\nu )}_{~ \delta}
\partial_{\alpha} \partial_{\gamma} \!-\! \delta^{(\mu}_{~~ \delta}
\delta^{\nu )}_{~\alpha} \partial_{\gamma} \partial_{\beta} \Bigr)
\; , \qquad \label{calD6} \\
\mathcal{D}_{\beta\delta}^{~~~ \mu\nu} & \equiv &
\eta^{\alpha\gamma} \mathcal{D}_{\alpha\beta\gamma\delta}^{~~~~~
\mu\nu} = -\frac12 \Bigl( \eta^{\mu\nu} \partial_{\beta}
\partial_{\delta} \!-\! 2 \partial^{(\mu} \delta^{\nu)}_{~ (\beta}
\partial_{\delta)} \!+\! \delta^{(\mu}_{~~ \beta} \delta^{\nu )}_{~
\delta} \partial^2 \Bigr) \; , \label{calD4} \qquad \\
\mathcal{D}^{\mu\nu} & \equiv & \eta^{\alpha\gamma}
\eta^{\beta\delta} \mathcal{D}_{\alpha\beta\gamma\delta}^{~~~~~
\mu\nu} = \partial^{\mu} \partial^{\nu} \!-\! \eta^{\mu\nu}
\partial^2 \; . \qquad \label{calD2}
\end{eqnarray}

Because the linearized Riemann tensor is invariant under linearized
gauge transformations ($\delta h_{\mu\nu} = -\partial_{\mu}
\xi_{\nu} - \partial_{\nu} \xi_{\mu}$) the operator
$\mathcal{D}_{\alpha\beta\gamma\delta}^{~~~~~ \mu\nu}$ and all its
traces are transverse on the indices $\mu$ and $\nu$. We also know
that the linearized Weyl tensor vanishes for a conformal graviton
field ($h_{\mu\nu}(x) = \eta_{\mu\nu} \Omega^2(x)$), all of which
implies that the operator
$\mathcal{C}_{\alpha\beta\gamma\delta}^{~~~~~ \mu\nu}$ obeys two
important identities,
\begin{equation}
\mathcal{C}_{\alpha\beta\gamma\delta}^{~~~~~ \mu\nu} \times
\eta_{\mu\nu} = 0 \qquad , \qquad
\mathcal{C}_{\alpha\beta\gamma\delta}^{~~~~~ \mu\nu} \times
\partial_{\mu} = 0 \; . \label{transtrace}
\end{equation}
These identities mean we can define suitable transverse-traceless
projectors by contracting
$\mathcal{C}_{\alpha\beta\gamma\delta}^{~~~~~ \mu\nu}(x)$ times
$\mathcal{C}_{\kappa\lambda\theta\phi}^{~~~~~ \rho\sigma}(x')$ into
any reflection invariant 8-index tensor. The choice made in the de
Sitter invariant construction \cite{PW1} was four products of the de
Sitter invariant bi-tensor $D^{\alpha} {D'}^{\kappa} y(x;x')$, but
that accounts for a large part of the complexity of the resulting
representation. A far simpler --- but noninvariant ---
representation will result from using products of
$\eta^{\alpha\kappa}$ and $\overline{\eta}^{\alpha\kappa} \equiv
\eta^{\alpha\kappa} + \delta^{\alpha}_{~0} \delta^{\kappa}_{~0}$,
\begin{eqnarray}
\mathcal{F}^{\mu\nu\rho\sigma} & \equiv &
\mathcal{C}_{\alpha\beta\gamma\delta}^{~~~~~ \mu\nu}(x) \times
\mathcal{C}_{\kappa\lambda\theta\phi}^{~~~~~ \rho\sigma}(x') \times
\eta^{\alpha\kappa} \eta^{\beta\lambda} \eta^{\gamma\theta}
\eta^{\delta\phi} \; , \label{calF} \\
\mathcal{G}^{\mu\nu\rho\sigma} & \equiv &
\mathcal{C}_{\alpha\beta\gamma\delta}^{~~~~~ \mu\nu}(x) \times
\mathcal{C}_{\kappa\lambda\theta\phi}^{~~~~~ \rho\sigma}(x') \times
\overline{\eta}^{\alpha\kappa} \overline{\eta}^{\beta\lambda}
\overline{\eta}^{\gamma\theta} \overline{\eta}^{\delta\phi} \; .
\label{calG}
\end{eqnarray}
Explicit expressions for these operators are given in the Appendix.

\section{Structure Functions for a MMC Scalar}\label{structure}

Actual computations of the graviton self-energy will initially take
the form of linear combinations of the basis tensors. The purpose of
this section is first to explain generally how to reduce this
initial primitive result to our form (\ref{ourform}). We then apply
this technique to work out the structure functions for the one loop
contribution from a massless, minimally coupled scalar \cite{PW1}.

\subsection{Finding the structure functions
generally}\label{general}

Suppose a primitive result for the graviton self-energy
$-i[\mbox{}^{\mu\nu} \Sigma^{\rho\sigma}](x;x')$ is known. Because
this primitive form can be expressed in the form (\ref{ourform}) we
can reconstruct the four structure functions by picking off
particularly simple tensor components. The procedure is first to
trace on one index group, which makes the spin two contributions
drop out, and then derive two linearly independent equations to
reconstruct $F_0(x;x')$ and $G_0(x;x')$. We then derive two linearly
independent equations to reconstruct $F_2(x;x')$ and $G_2(x;x')$. In
what follows we will always assume $i \neq j \neq k \neq i$.

Tracing on $\rho$ and $\sigma$ gives,
\begin{equation}
-i \Bigl[\mbox{}^{\mu\nu} \Sigma^{\rho\sigma}\Bigr](x;x') \times
\eta_{\rho\sigma} = \mathcal{F}^{\mu\nu} \Biggl( {a'}^D \mathcal{F}'
\Bigl[ \frac{F_0(x;x')}{{a'}^{D-2}} \Bigr] \Biggr) +
\mathcal{G}^{\mu\nu} \Biggl( {a'}^D \mathcal{G}' \Bigl[
\frac{G_0(x;x')}{ {a'}^{D-2} } \Bigr] \Biggr) \; . \label{step1}
\end{equation}
Our equations derive from the simple forms attained by the scalar
projectors for $\mu = 0$, $\nu = i$,
\begin{eqnarray}
\mathcal{F}^{0i} & = & a^{D-2} \times \Bigl[ -\partial_0 \!+\! a
H\Bigr] \partial_i \times a^{-(D-2)} \; , \label{F0i} \\
\mathcal{G}^{0i} & = & a^{D-2} \times \Bigl[ (D \!-\! 2) a H \Bigr]
\partial_i \times a^{-(D-2)} \; , \label{G0i}
\end{eqnarray}
and for $\mu = j$, $\nu = k\neq j$,
\begin{equation}
\mathcal{F}^{jk} = a^{D-2} \times \partial_j \partial_k \times
a^{-(D-2)} \qquad , \qquad \mathcal{G}^{jk} = a^{D-2} \times
\partial_j \partial_k \times a^{-(D-2)} \; . \label{FGjk}
\end{equation}
By homogeneity and isotropy, these same index combinations for the
self-energy must produce the same spatial derivatives. Because the
self-energy is $(a a')^2$ times a contravariant bi-tensor density,
and $\eta_{\rho\sigma} = g_{\rho\sigma}(x') \times {a'}^{-2}$, it
also makes sense to extract a factor of $a^{D-2} {a'}^D$,
\begin{eqnarray}
-i \Bigl[ \mbox{}^{0i} \Sigma^{\rho\sigma} \Bigr](x;x') \times
\eta_{\rho\sigma} & \equiv & a^{D-2} {a'}^D \partial_i S_1(x;x') \;
, \label{0isource} \\
-i \Bigl[ \mbox{}^{jk} \Sigma^{\rho\sigma} \Bigr](x;x') \times
\eta_{\rho\sigma} & \equiv & a^{D-2} {a'}^D \partial_j \partial_k
S_2(x;x') \; . \label{jksource}
\end{eqnarray}
Comparing relations (\ref{F0i}-\ref{G0i}) with (\ref{0isource})
implies,
\begin{equation}
\Bigl[ -\partial_0 \!+\! a H \Bigr] \Biggl( \mathcal{F}' \Bigl[
\frac{F_0(x;x')}{(a a')^{D-2} } \Bigr] \Biggr) + (D\!-\! 2) a H
\Biggl( \mathcal{G}' \Bigl[ \frac{G_0(x;x')}{(a a')^{D-2} } \Bigr]
\Biggr) = S_1(x;x') \; . \label{eqn1}
\end{equation}
The second independent equation comes from comparing (\ref{FGjk})
with (\ref{jksource}),
\begin{equation}
\mathcal{F}' \Bigl[ \frac{F_0(x;x')}{(a a')^{D-2} } \Bigr] +
\mathcal{G}' \Bigl[ \frac{G_0(x;x')}{(a a')^{D-2} } \Bigr] =
S_2(x;x') \; . \label{eqn2}
\end{equation}

Given the two source functions $S_1(x;x')$ and $S_2(x;x')$, we can
obtain an equation for $F_0(x;x')$ by subtracting $(D-2) a H$ times
(\ref{eqn2}) from (\ref{eqn1}),
\begin{equation}
\Bigl[\partial_0 + (D \!-\! 3) a H \Bigr] \Biggl( \mathcal{F}'
\Bigl[ \frac{F_0(x;x')}{(a a')^{D-2} } \Bigr] \Biggr) = -S_1(x;x') +
(D \!-\! 2) a H S_2(x;x') \; . \label{F0eqn}
\end{equation}
The solution can be expressed as an indefinite integral,
\begin{equation}
\mathcal{F}' \Bigl[ \frac{F_0(x;x')}{(a a')^{D-2} } \Bigr] =
a^{-(D-3)} \!\! \int \!\! d\eta \, a^{D-3} \Bigl[-S_1(x;x') + (D
\!-\!2) a H S_2(x;x') \Bigr] \; . \label{F0sol}
\end{equation}
The comparable relation for $G_0(x;x')$ comes from subtracting
(\ref{F0sol}) from (\ref{eqn2}),
\begin{equation}
\mathcal{G}' \Bigl[ \frac{G_0(x;x')}{(a a')^{D-2} } \Bigr] =
S_2(x;x') + a^{-(D-3)} \!\! \int \!\! d\eta \, a^{D-3}
\Bigl[S_1(x;x') - (D \!-\!2) a H S_2(x;x') \Bigr] \; . \label{G0sol}
\end{equation}

We can recover the structure function $F_0(x;x')$ from expression
(\ref{F0sol}) by employing the Green's function for $\mathcal{F}'$,
\begin{equation}
\mathcal{F}' = -(D \!-\! 1) \Bigl[ \square' \!+\! D H^2 \Bigr] \; .
\end{equation}
This Greens' function is proportional to the scalar propagator for a
tachyonic mass of $M^2 = -D H^2$, and its specialization to a de
Sitter invariant source was derived in \cite{PW1}. The structure
function $G_0(x;x')$ comes from integrating expression (\ref{G0sol})
against the Green's function for $\mathcal{G}'$,
\begin{equation}
\mathcal{G}' = -\frac{(D \!-\! 2)}{ {a'}^2} \Bigl[ {\nabla'}^2 + (D
\!-\! 1) a' H \partial'_0 + D(D\!-\!1) {a'}^2 H^2
\Bigr] \; .
\end{equation}

The key to determining the spin two structure functions is a set of
four identities for the projectors $\mathcal{F}^{0ijk}$ (with $i \neq
j \neq k \neq i$) which can be derived for (after using homogeneity
to reflect spatial derivatives $\partial'_i = -\partial_i$) from the
explicit forms given in the Appendix,
\begin{eqnarray}
\mathcal{F}^{0ijk} & = & \frac{ (D \!-\! 3)}{(D \!-\! 1) (D \!-\!
2)} \Bigl[ (D \!-\! 1) \partial'_0 + \partial_0 \Bigr] \partial_i
\partial_j \partial_k \; , \label{F0ijk} \\
\mathcal{F}^{jk0i} & = & \frac{ (D \!-\! 3)}{(D \!-\! 1) (D \!-\!
2)} \Bigl[ - \partial'_0 - (D \!-\! 1) \partial_0 \Bigr]
\partial_i \partial_j \partial_k \; , \qquad \label{Fjk0i} \\
\mathcal{G}^{0ijk} & = & \frac{ (D \!-\! 3)}{(D \!-\! 1) (D \!-\!
2)^2} \times \partial_0 \partial_i \partial_j \partial_k \; ,
\label{G0ijk} \\
\mathcal{G}^{jk0i} & = & \frac{ (D \!-\! 3)}{(D \!-\! 1) (D \!-\!
2)^2} \times - \partial'_0 \partial_i \partial_j \partial_k \; .
\label{Gjk0i}
\end{eqnarray}
Assuming the spin zero structure functions are known we can
reconstruct the spin two structure functions from sums and
differences of the $0ijk$ and $jk0i$ components. From homogeneity
and isotropy, and a judicious guess for the scale factors, we can
express these components as,
\begin{eqnarray}
-i \Bigl[\mbox{}^{0i} \Sigma^{jk}\Bigr](x;x') & = & (a a')^{D-2}
\partial_i \partial_j \partial_k S_3(x;x') \; , \label{S3} \\
- i \Bigl[ \mbox{}^{jk} \Sigma^{0i} \Bigr](x;x') & = & (a a')^{D-2}
\partial_i \partial_j \partial_k S_4(x;x') \; . \label{S4}
\end{eqnarray}
Combining these relations with (\ref{F0ijk}-\ref{Gjk0i}), and
expressions (\ref{F0i}-\ref{FGjk}) allows us to derive first order
differential equations for the spin two structure functions,
\begin{eqnarray}
\lefteqn{ \frac{(D \!-\!3) (\partial'_0 \!-\! \partial_0) }{(D \!-\!
1) (D\!-\! 2)^2} \, \Bigl[ (D \!-\! 2)^2 F_2(x;x') \!-\!
G_2(x;'x)\Bigr] = (a a')^{D-2} \Bigl[ S_3 \!+\! S_4 \Bigr] }
\nonumber \\
& & \hspace{2.2cm} - \Bigl[ \partial'_0 \!-\! \partial_0 \!+\! (D
\!-\! 1) (a \!-\! a') H\Bigr] F_0 - (D \!-\! 2) (a \!-\! a') H
G_0 \; , \label{eqn3} \qquad \\
\lefteqn{ \frac{(D \!-\!3) (\partial'_0 \!+\! \partial_0) }{(D \!-\!
1) (D\!-\! 2)^2} \, \Bigl[ D (D \!-\! 2) F_2(x;x') \!+\!
G_2(x;'x)\Bigr] = (a a')^{D-2} \Bigl[ S_3 \!-\! S_4 \Bigr] }
\nonumber \\
& & \hspace{2.2cm} + \Bigl[ \partial'_0 \!+\! \partial_0 \!-\! (D
\!-\! 1) (a \!+\! a') H\Bigr] F_0 - (D \!-\! 2) (a \!+\! a') H G_0
\; . \label{eqn4} \qquad
\end{eqnarray}
Because equations (\ref{eqn3}) and (\ref{eqn4}) determine different
linear combinations of $F_2(x;x')$ and $G_2(x;x')$ we can recover
both of the spin two structure functions.

\subsection{Primitive contribution from a MMC
scalar}\label{primitive}

The contribution to the graviton self-energy from a loop of
massless, minimally coupled (MMC) scalars can be expressed as a
linear combination of the de Sitter invariant basis tensors
\cite{PW1,PNRV},
\begin{eqnarray}
\lefteqn{-i \Bigl[\mbox{}^{\mu\nu} \Sigma^{\rho\sigma} \Bigr] = (a
a')^{D+2} \Biggl\{ D^{\mu} {D'}^{(\rho} y {D'}^{\sigma)} D^{\nu} y
\, \alpha + D^{(\mu} y D^{\nu)} {D'}^{(\rho} y {D'}^{\sigma)} y \,
\beta } \nonumber \\
& & \hspace{0cm} + D^{\mu} y D^{\nu} y {D'}^{\rho} y {D'}^{\sigma} y
\, \gamma \!+\! H^4 g^{\mu\nu} {g'}^{\rho\sigma} \delta \!+\! H^2
\Bigl[ g^{\mu\nu} {D'}^{\rho}y {D'}^{\sigma} y \!+\! (\rm refl.)
\Bigr] \epsilon \Biggr\} . \qquad \label{Sohyun}
\end{eqnarray}
The combination coefficients are functions of the de Sitter length
function $y(x;x')$ and derivatives of the function $A(y)$
\cite{PW1},
\begin{equation}
\alpha(y) = -\frac{\kappa^2}2 [A'(y)]^2 \quad , \quad \beta(y) =
-\kappa^2 A'(y) A''(y) \quad , \quad \gamma(y) = -\frac{\kappa^2}2
[A''(y)]^2 \; , \label{alpha}
\end{equation}
\begin{equation}
\delta(y) = -\frac{\kappa^2}8 \Biggl[ (4 y \!-\! y^2)^2 (A'')^2
\!+\! 2 (2 \!-\! y) (4 y \!-\! y^2) A' A'' \!+\! \Bigl[ 4(D\!-\!4)
\!-\! (4y \!-\! y^2)\Bigr] (A')^2 \Biggr] , \label{delta}
\end{equation}
\begin{equation}
\epsilon(y) = \frac{\kappa^2}4 \Biggl[ (4 y \!-\! y^2) [A''(y)]^2
\!+\! 2 (2 \!-\! y) A'(y) A''(y) \!-\! [A'(y)]^2 \Biggr] .
\label{gamma}
\end{equation}

The function $A(y)$ is the de Sitter invariant part of the MMC
scalar propagator $i\Delta(x;x')=A(y)+k\ln(aa')$~\cite{OW},
\begin{eqnarray}
\lefteqn{A(y) = \frac{H^{D-2}}{(4\pi)^{\frac{D}2}} \Biggl\{
\frac{\Gamma(\frac{D}2)}{\frac{D}2 \!-\! 1} \Bigl( \frac4{y}
\Bigr)^{\frac{D}2 -1} + \frac{ \Gamma(\frac{D}2 \!+\! 1)}{\frac{D}2
\!-\! 2} \Bigl(\frac4{y}\Bigr)^{\frac{D}2 -2} + {\rm constant} }
\nonumber \\
& & \hspace{0cm} + \sum_{n=1}^{\infty} \Biggr[ \frac1{n}
\frac{\Gamma(n \!+\! D \!-\! 1)}{\Gamma(n \!+\! \frac{D}2)}
\Bigl(\frac{y}4 \Bigr)^n - \frac1{n \!-\! \frac{D}2 \!+\! 2}
\frac{\Gamma(n \!+\! \frac{D}2 \!+\! 1)}{\Gamma(n \!+\! 2)}
\Bigl(\frac{y}4 \Bigr)^{n - \frac{D}2 +2} \!\Biggr] \Biggr\} ,
\qquad \label{A(y)}
\end{eqnarray}
Note that the infinite series on the second line of (\ref{A(y)})
vanishes in $D=4$, so these terms only need to be retained when they
multiply a divergent term. It is also worth noting that $A(y)$ obeys
the equation,
\begin{equation}
(4 y \!-\! y^2) A''(y) + D (2 \!-\! y) A'(y) = \frac{ H^{D-2}}{(4
\pi)^{\frac{D}2}} \frac{ \Gamma(D)}{\Gamma(\frac{D}2 )}
\equiv (D\!-\!1)k \; .
\end{equation}

\subsection{Spin zero structure functions for a MMC
scalar}\label{F0G0}

The trace of expression (\ref{Sohyun}) produces one term
proportional to $g^{\mu\nu}$ and another proportional to $D^{\mu} y
D^{\nu}y$,
\begin{eqnarray}
\lefteqn{-i \Bigl[\mbox{}^{\mu\nu} \Sigma^{\rho\sigma} \Bigr](x;x')
\times \eta_{\rho\sigma} = a^2 H^2 (a a')^{D} \Biggl\{ H^2
g^{\mu\nu} \Bigl[4 \alpha + D \delta + (4 y \!-\! y^2) \epsilon
\Bigr] } \nonumber \\
& & \hspace{3.5cm} + D^{\mu} y D^{\nu} y \Bigl[ -\alpha + (2 \!-\!
y) \beta \!+\! (4 y \!-\! y^2) \gamma + D \epsilon \Bigr] \Biggr\} .
\qquad
\end{eqnarray}
Now recall from (\ref{basis1}) that the components we need of
$D^{\mu} y D^{\nu} y$ are,
\begin{equation}
D^0 y D^i y = a^{-4} \times -a H \Bigl(y \!-\! 2 \!+\! 2
\frac{a'}{a} \Bigr) \partial_i y \quad , \quad D^j y D^k y = a^{-4}
\times \partial_j y \partial_k y \; . \label{DyDy}
\end{equation}
The two scalar sources follow from comparison with expressions
(\ref{0isource}-\ref{jksource}),
\begin{eqnarray}
S_1(x;x') & = & -a H^3 I\Bigl[ (y \!-\! 2) F''\Bigr] - 2 a' H^3
F'(y) = H^2 \Bigl( -\partial_0 \!+\! a H\Bigr) F(y) \; , \label{S1}
\qquad \\
S_2(x;x') & = & H^2 F(y) \; , \label{S2}
\end{eqnarray}
where the function $F(y)$ is a double indefinite
integral,\footnote{We define the indefinite integral of a function
$f(y)$ as $I[f](y) \equiv \int^y dz f(z)$.}
\begin{equation}
F(y) \equiv I^2\Bigl[ -\alpha + (2 \!-\! y) \beta + (4 y \!-\! y^2)
\gamma + D \epsilon \Bigr] \; . \label{Fdef}
\end{equation}

Substituting (\ref{S1}-\ref{S2}) into expressions (\ref{F0sol}) and
(\ref{G0sol}) gives the spin zero structure functions,
\begin{equation}
F_0(x;x') = -\frac{(a a')^{D-2}}{D \!-\! 1} \Bigl(
\frac{H^2}{\square \!+\! D H^2} \Bigr) F(y) \quad , \quad
G_0(x;x') = 0 \; . \label{F0G0sol}
\end{equation}
In retrospect we can observe that the vanishing of $G_0(x;x')$ is a
consequence of the fact that our first spin zero term
$\mathcal{F}^{\mu\nu} \times {\mathcal{F}'}^{\rho\sigma}
[F_0(x;x')]$ is just a conformal transformation of the single spin
zero contribution in the de Sitter invariant construction
\cite{PW1}. The second spin zero structure function $G_0(x;x')$ must
vanish whenever the graviton self-energy is de Sitter invariant.

In principle, we could read off the fully renormalized result for
$F_0(x;x')$ as $(a a')^{D-2} \times \mathcal{F}_{1R}(y)$, from
equation (234) of \cite{PW1}. However, there are some subtleties to
inverting $\square + D H^2$ that were not previously understood, and
our current approach has the significant simplification of only
requiring a single inversion rather than two \cite{PW1}. We will
therefore carry out the derivation.

Substituting expression (\ref{A(y)}) into (\ref{alpha}-\ref{gamma}),
and then into (\ref{Fdef}) gives the following expansion for the
function $F(y)$,
\begin{equation}
F(y) = \frac{\kappa^2 H^{2D-4} \Gamma^2(\frac{D}2)}{(4 \pi)^{D}}
\Biggl\{ \frac{(D\!-\!2)^2}{16 (D \!-\!1)} \Bigl( \frac{4}{y}
\Bigr)^{D-1} \!\!\!\!\!\! + \frac{(D^3 \!-\! 5 D^2 \!+\! 6 D \!-\!
4)}{16 (D \!-\! 1)} \Bigl( \frac{4}{y}\Bigr)^{D-2} \!\!\!\!\!\! +
\dots \Biggr\} . \label{Fexp}
\end{equation}
The neglected terms in this and subsequent expansions have the
twin properties that they make integrable contributions to the
structure functions, and they vanish in $D=4$ dimensions. Inverting
$(\square/H^2 + D)$ on $F(y)$ amounts to solving the differential
equation,
\begin{equation}
\Bigl[ \frac{\square}{H^2} + D\Bigr] f(x;x') = F\Bigl(y(x;x')\Bigr)
\; . \label{feqn}
\end{equation}
The first step is to expand $f(x;x')$ so as to absorb the leading
terms of $F(y)$ in expression (\ref{Fexp}),
\begin{eqnarray}
\lefteqn{f(x;x') = \frac{\kappa^2 H^{2D-4} \Gamma^2(\frac{D}2)}{(
4 \pi)^{D}} \Biggl\{ \frac1{8 (D \!-\!1)} \Bigl( \frac{4}{y}\Bigr)^{D-2}
} \nonumber \\
& & \hspace{3cm} + \frac{D (D^2 \!-\! 5 D \!+\! 2)}{8 (D \!-\! 4)
(D \!-\! 3) (D \!-\! 1)} \Bigl( \frac{4}{y}\Bigr)^{D-3}
+ \Delta f_0(x;x') \Biggr\} , \qquad \label{f0exp}
\end{eqnarray}
where the remainder $\Delta f_0(x;x')$ obeys,
\begin{equation}
\Bigl[ \frac{\square}{H^2} + D\Bigr] \Delta f_0(x;x') = -\frac{3 (D\!-\!2)
D (D^2 \!-\! 5 D \!+\! 2)}{8 (D \!-\! 4) (D \!-\! 3) (D \!-\! 1)}
\Bigl( \frac{4}{y}\Bigr)^{D-3} \; . \label{Df0eqn}
\end{equation}

Note that we could set $D=4$ for the $(4/y)^{D-3}$ terms of
expressions (\ref{f0exp}) and (\ref{Df0eqn}) were it not for the
multiplicative factors of $1/(D-4)$. We can localize the divergence by
adding zero based on the identity,\footnote{In dimensional regularization
it is easy to show that one only gets delta functions from differentiating
$1/y^{\frac{D}2 + N}$, where $N = -1, 0, 1, 2 \dots$. Even terms which
become identical in four dimensions --- for example, $1/y^{D + N -2}$ ---
do not produce delta functions \cite{OW}.}
\begin{equation}
\Bigl[ \frac{\square}{H^2} + D\Bigr] \Bigl(\frac{4}{y} \Bigr)^{\frac{D}2 -1}
= \frac{D (D \!+\! 2)}{4} \Bigl( \frac{4}{y}\Bigr)^{\frac{D}2 -1} +
\frac{ (4\pi)^{\frac{D}2} \, i \delta^D(x \!-\! x')}{(H a)^D \Gamma(\frac{D}2
\!-\! 1)} \; . \label{zero}
\end{equation}
The revised expansion becomes,
\begin{eqnarray}
\lefteqn{f(x;x') = \frac{\kappa^2 H^{2D-4} \Gamma^2(\frac{D}2)}{(
4 \pi)^{D}} \Biggl\{ \frac1{8 (D \!-\!1)} \Bigl( \frac{4}{y}\Bigr)^{D-2}
} \nonumber \\
& & \hspace{1cm} + \frac{D (D^2 \!-\! 5 D \!+\! 2)}{8 (D \!-\! 4)
(D \!-\! 3) (D \!-\! 1)} \Bigl[\Bigl( \frac{4}{y}\Bigr)^{D-3} \!\!\!\!-
\Bigl( \frac{4}{y} \Bigr)^{\frac{D}2 -1} \Bigr] + \Delta f_1(x;x')
\Biggr\} , \qquad \label{f1exp}
\end{eqnarray}
where the new remainder $\Delta f_1(x;x')$ obeys,
\begin{eqnarray}
\lefteqn{\Bigl[ \frac{\square}{H^2} + D\Bigr] \Delta f_1(x;x') = \frac{D
(D^2 \!-\! 5 D \!+\! 2)}{8 (D \!-\! 4) (D \!-\! 3) (D \!-\! 1)} \times
\frac{ (4\pi)^{\frac{D}2} \, i \delta^D(x \!-\! x')}{(H a)^D \Gamma(\frac{D}2
\!-\! 1)} } \nonumber \\
& & \hspace{2cm} - \frac{3 (D\!-\!2) D (D^2 \!-\! 5 D \!+\! 2)}{8
(D \!-\! 4) (D \!-\! 3) (D \!-\! 1)} \Bigl[ \Bigl( \frac{4}{y}\Bigr)^{D-3}
\!\!\!\!\!- \frac{D (D \!+\! 2)}{12 (D \!-\! 2)} \Bigl( \frac{4}{y}
\Bigr)^{\frac{D}2 - 1} \Bigr] \; . \qquad \label{Df1eqn}
\end{eqnarray}
The term proportional to the delta function in expression (\ref{Df1eqn})
can be absorbed into the counterterm $\Delta \mathcal{L}_3 = c_3 H^2 [R -
(D-1)(D-2) H^2] \sqrt{-g}$ of Ref.~\cite{PW1}, see especially equations
(115) and (228) of that work. When this is done we can set $D=4$ in all
but the first term of (\ref{f1exp}). The resulting, partially renormalized
expansion is,
\begin{equation}
f(x;x') = \frac{\kappa^2 H^{2D-4} \Gamma^2(\frac{D}2)}{(
4 \pi)^{D}} \Biggl\{ \frac1{8 (D \!-\!1)} \Bigl( \frac{4}{y}\Bigr)^{D-2}
\!\!\!\! + \frac{2 \ln(\frac{y}4)}{3 y} + \Delta f_2(x;x') \Biggr\} ,
\label{f2exp}
\end{equation}
where the renormalized remainder $\Delta f_2(x;x')$ obeys,
\begin{equation}
\Bigl[ \frac{\square}{H^2} + 4\Bigr] \Delta f_2(x;x') = -\Bigl[\frac{4
\ln(\frac{y}4) \!-\! \frac23}{y} \Bigr] \equiv \Delta F_2(y) \; .
\label{Df2eqn}
\end{equation}

The next step is to derive a formal solution to (\ref{Df2eqn}) using the
de Sitter invariant Green's function that can be constructed from the
two homogeneous solutions \cite{PW1},
\begin{eqnarray}
\phi_1(y) & = & 2 - y \; , \label{phi1} \\
\phi_2(y) & = & -\frac{2}{y} - \frac{2}{4 \!-\! y} + \frac32 (2 \!-\! y)
\Bigl[ \ln\Bigl(\frac{y}4\Bigr) - \ln\Bigl(1 \!-\! \frac{y}4\Bigr)
\Bigr] + 6 \; . \qquad \label{phi2}
\end{eqnarray}
Although (\ref{phi1}) obeys $(\square + 4 H^2) \phi_1(y) = 0$, acting
$(\square + 4 H^2)$ on expression (\ref{phi2}) actually produces delta
functions at the origin and at the antipodal point. (This is the
subtlety that was not understood in the original construction
\cite{PW1}.) We can nonetheless derive a homogeneous and isotropic
solution for $\Delta f_2(x;x')$ by a process of employing the formal
Green's function,
\begin{equation}
\mathcal{G}(y;y') = \theta(y \!-\! y') \Bigl[ \phi_2(y) \phi_1(y') \!-\!
\phi_1(y) \phi_2(y')\Bigr] \mathcal{W}(y') \;\; , \;\; \mathcal{W}(y)
\equiv \frac{(4y \!-\! y^2)}{64} \; , \label{Gfunc}
\end{equation}
and then subtracting off the unwanted pole terms.

The formal, de Sitter invariant solution to (\ref{Df2eqn}) can be
expressed using the indefinite integral operation $I[\dots]$,
\begin{eqnarray}
\lefteqn{\phi_2(y) I\Bigl[\phi_1 \mathcal{W} \Delta F_2\Bigr](y) -
\phi_1(y) I\Bigl[\phi_2 \mathcal{W} \Delta F_2\Bigr](y) =
\frac{-2}{4 \!-\! y}
- \frac{29}3 + \frac{41}6 y
} \nonumber \\
& & \hspace{1.5cm} + \biggl[ \frac{\frac23}{4 \!-\! y} + 1 - \frac32 y
\biggr] \ln\Bigl( \frac{y}{4} \Bigr) - (2 \!-\! y) \biggl[\frac32
\ln\Bigl(1 \!-\! \frac{y}{4}\Bigr) \!+\! \frac12 \Psi(y)\biggr] \; ,
\qquad \label{sol1}
\end{eqnarray}
where the function $\Psi(y)$ will appear in all the structure
functions,
\begin{equation}
\Psi(y) \equiv \frac12 \ln^2\Bigl( \frac{y}4\Bigr) - \ln\Bigl(1 \!-\!
\frac{y}4\Bigr) \ln\Bigl( \frac{y}4\Bigr) - {\rm Li}_2\Bigl(\frac{y}4
\Bigr) \; . \label{Psidef}
\end{equation}
Expression (\ref{sol1}) is only a formal solution to (\ref{Df2eqn})
because of the pole it has at the antipodal point of $y = 4$. We can
eliminate this pole by adding $\frac{41}6 \phi_1(y) -\phi_2(y)$,
\begin{equation}
\Delta f_2(x;x') = \frac{2}{y} - 2
+ \biggl[ \frac{\frac23}{4 \!-\! y} \!-\! 2\biggr]
\ln\Bigl( \frac{y}{4} \Bigr) - \Bigl(\frac{2 \!-\! y}2\Bigr) 
\Psi(y) \; . \label{Df2}
\end{equation}

Substituting expressions (\ref{Df2}) and (\ref{f2exp}) into
(\ref{F0G0sol}) allows us to at length express the scalar structure
function as $F_0(x;x') = (a a')^{D-2} \Phi(y)$ where,
\begin{eqnarray}
\lefteqn{\Phi(y) = \frac{\kappa^2 H^{2D-4} \Gamma^2(\frac{D}2)}{
(4 \pi)^{D}} \Biggl\{ -\frac1{8 (D \!-\!1)^2} \Bigl( \frac{4}{y}
\Bigr)^{D-2} } \nonumber \\
& & \hspace{2.5cm} -\frac{2}{3y} -\frac29 \biggl[ \frac1{y} \!-\! 
3 \!+\! \frac1{4 \!-\! y} \biggr] \ln\Bigl(\frac{y}4\Bigr) + 
\frac23 + \Bigl(\frac{2 \!-\! y}{6} \Bigr) \Psi(y) \Biggr\} . \qquad
\label{phidef}
\end{eqnarray}
The most singular part of $F_0(x;x')$ can be partially integrated
to isolate the remaining ultraviolet divergence,
\begin{eqnarray}
\lefteqn{ \frac{\kappa^2 (H^2 a a')^{D-2} \Gamma^2(\frac{D}2)}{(4
\pi)^D} \times -\frac1{8 (D\!-\! 1)^2} \Bigl( \frac{4}{y}
\Bigr)^{D-2} } \nonumber \\
& & = \frac{\kappa^2 \Gamma^2(\frac{D}2)}{ 16 \pi^D} \times
-\frac{\partial^2}{16 (D \!-\! 4) (D\!-\! 3) (D\!-\! 1)^2}
\Bigl[\frac1{\Delta x^{2D - 6}} \Bigr] \; , \qquad \\
& & = \frac{\kappa^2 \Gamma^2(\frac{D}2)}{16 \pi^D} \frac{-1}{16
(D \!-\! 4) (D \!-\! 3) (D\!-\! 1)^2} \nonumber \\
& & \hspace{2.5cm} \times \Biggl\{ \partial^2 \Bigl[
\frac1{\Delta x^{2D-6}} - \frac{ \mu^{D-4}}{\Delta x^{D-2}} \Bigr]
+ \frac{4 \pi^{\frac{D}2} \mu^{D-4} i \delta^D(x \!-\! x')}{
\Gamma( \frac{D}2 \!-\! 1)} \Biggr\} \; , \qquad \\
& & \longrightarrow \frac{\kappa^2}{ (4\pi)^4} \frac{\partial^2}{18}
\Bigl[ \frac{\ln(\mu^2 \Delta x^2)}{\Delta x^2} \Bigr] -
\frac{\kappa^2 \mu^{D-4} \Gamma(\frac{D}2)}{128 \pi^{\frac{D}2}}
\frac{(D \!-\! 2) \, i \delta^D(x \!-\! x')}{(D\!-\!4)
(D \!-\! 3) (D \!-\! 1)^2} \; . \qquad \label{renform}
\end{eqnarray}
The divergence in expression (\ref{renform}) can be absorbed into
the counterterm $\Delta \mathcal{L}_1 = c_1 [R - D (D-1) H^2]^2
\sqrt{-g}$ of Ref. \cite{PW1}, see especially equations (113) and
(227) of that work.\footnote{Complete agreement with Ref.
\cite{PW1} requires our renormalization scale to be $\mu =
\frac12 H$.} Our final result for the renormalized scalar
structure function is therefore,
\begin{eqnarray}
\lefteqn{F_{0R}(x;x') = \frac{\kappa^2 (aa'H^2)^2}{2304 \pi^4}
\Biggl\{\frac{\partial^2}{2 (aa'H^2)^2} \Bigl[ \frac{
\ln(\mu^2 \Delta x^2)}{\Delta x^2} \Bigr] - \frac{6}{y}+ 6} 
\nonumber \\
& & \hspace{4cm} + \Bigl[ -\frac2{y} \!+\! 6 \!-\! 
\frac2{4 \!-\! y} \Bigr] \ln\Bigl(\frac{y}4\Bigr) + \frac32 
(2 \!-\! y) \Psi(y) \Biggr\} . \qquad \label{F0R}
\end{eqnarray}

\subsection{Spin two structure functions for a MMC
scalar}\label{F2G2}

In addition to the identities (\ref{DyDy}) and their reflections,
extracting the sources $S_3(x;x')$ and $S_4(x;x')$ from the $0ijk$
and $jk0i$ components requires an identity we can infer from
(\ref{basis2}) for the mixed derivative,
\begin{equation}
D^0 {D'}^j y = (a a')^{-2} \times a H \partial_j y \qquad , \qquad
D^i {D'}^0 y = (a a')^{-2} \times -a' H \partial_i y \; .
\label{DDy}
\end{equation}
Applying these identities to the desired components of
(\ref{Sohyun}) gives,
\begin{eqnarray}
\lefteqn{-i \Bigl[ \mbox{}^{0i} \Sigma^{jk}\Bigr](x;x') = (a
a')^{D+2} \Bigl\{ \frac12 D^i y D^0 {D'}^{(j} y {D'}^{k)} y \, \beta
+ D^0y D^i y {D'}^{j} y {D'}^{k} y \, \gamma \Bigr\} \; , }
\nonumber \\
& & \hspace{.5cm} = (a a')^{D-2} \partial_i \partial_j \partial_k
\Bigl\{ -\frac12 a H I^3[\beta] - a H I^3[(y \!-\! 2) \gamma] - 2 a'
H I^3[\gamma] \Bigr\} \; , \qquad \\
\lefteqn{-i \Bigl[ \mbox{}^{jk} \Sigma^{0i}\Bigr](x;x') = (a
a')^{D+2} \Bigl\{ \frac12 D^{(j} y D^{k)} {D'}^{0} y {D'}^{i} y \,
\beta + D^j y D^k y {D'}^{0} y {D'}^{i} y \, \gamma \Bigr\} \; , }
\nonumber \\
& & \hspace{.5cm} = (a a')^{D-2} \partial_i \partial_j \partial_k
\Bigl\{ \frac12 a' H I^3[\beta] + a' H I^3[(y \!-\! 2) \gamma] + 2 a
H I^3[\gamma] \Bigr\} \; . \qquad
\end{eqnarray}
Use of the partial integration identity $I^3[(y-2)\gamma] = (y-2)
I^3[\gamma] - 3 I^4[\gamma]$ and comparison with expressions
(\ref{S3}-\ref{S4}) implies,
\begin{eqnarray}
S_3(x;x') & = & -\frac12 a H I^3[\beta] + \Bigl( -\partial_0 \!+\! 3
a H\Bigr) I^4[\gamma] \; , \label{ParkS3} \\
S_4(x;x') & = & +\frac12 a' H I^3[\beta] + \Bigl( \partial_0' \!-\!
3 a' H \Bigr) I^4[\gamma] \; . \label{ParkS4}
\end{eqnarray}

Substituting expressions (\ref{ParkS3}-\ref{ParkS4}) into
(\ref{eqn3}-\ref{eqn4}), and setting the spin zero structure
functions to $F_0(x;x') \equiv (a a')^{D-2} \times \Phi(y)$ and
$G_0(x;x') = 0$, gives equations for $F_2(x;x')$ and $G_2(x;x')$,
\begin{eqnarray}
\lefteqn{ (\partial'_0 \!-\! \partial_0) \Biggl\{ \Bigl( \frac{D
\!-\! 3}{D \!-\! 1} \Bigr) \, F_2 - \frac{(D \!-\! 3)\, G_2}{(D
\!-\! 1) (D \!-\! 2)^2} - (a a')^{D-2} I^4[\gamma] + F_0 \Biggr\} }
\nonumber \\
& & \hspace{1cm} = (a \!-\! a') H \times (a a')^{D-2}
\Bigl\{-\frac12 I^3[\beta] + (D\!+\!1) I^4[\gamma] - (D \!-\! 1)
\Phi \Bigr\} \; , \label{eqnA} \qquad \\
\lefteqn{ (\partial'_0 \!+\! \partial_0) \Biggl\{ \frac{D (D \!-\!
3)}{(D \!-\! 1) (D \!-\! 2)} \, F_2 + \frac{(D \!-\! 3)\, G_2}{(D
\!-\! 1) (D \!-\! 2)^2} + (a a')^{D-2} I^4[\gamma] - F_0 \Biggr\} }
\nonumber \\
& & \hspace{1cm} = (a \!+\! a') H \times (a a')^{D-2}
\Bigl\{-\frac12 I^3[\beta] + (D\!+\!1) I^4[\gamma] - (D \!-\! 1)
\Phi \Bigr\} \; , \label{eqnB} \qquad
\end{eqnarray}
To solve equation (\ref{eqnA}), consider acting $\partial'_0 -
\partial_0$ on $(a a')^{D-2} \times f(y)$,
\begin{equation}
(\partial'_0 \!-\! \partial_0) \Bigl[ (a a')^{D-2} f(y) \Bigr] = (a
\!-\! a') H (a a')^{D-2} \Bigl\{ (4 \!-\! y) f'(y) - (D
\!-\! 2) f(y) \Bigr\} \; .
\end{equation}
The solution to (\ref{eqnA}) can therefore be expressed as an
indefinite integral,
\begin{eqnarray}
\lefteqn{\Bigl( \frac{D \!-\! 3}{D \!-\! 1} \Bigr) \, F_2 - \frac{(D
\!-\! 3)\, G_2}{(D \!-\! 1) (D \!-\! 2)^2} - (a a')^{D-2}
I^4[\gamma] + F_0 = \frac{(a a')^{D-2}}{(4 \!-\! y)^{D-2}}
} \nonumber \\
& & \hspace{.3cm} \times \Biggl\{
I\Biggl[ (4 \!-\! y)^{D-3} \Bigl\{-\frac12 I^3[\beta] + (D\!+\!1)
I^4[\gamma] - (D \!-\! 1) \Phi \Bigr\} \Biggr] + K_1 \Biggr\} ,
\qquad \label{finalA}
\end{eqnarray}
where the integration constant $K_1$ is chosen so that there is no
singularity at the antipdal point $y = 4$. Of course the same
considerations apply for equation (\ref{eqnB}),
\begin{equation}
(\partial'_0 \!+\! \partial_0) \Bigl[ (a a')^{D-2} f(y) \Bigr] = (a
\!+\! a') H (a a')^{D-2} \Bigl\{ y f'(y) + (D \!-\! 2) f(y)
\Bigr\} \; .
\end{equation}
The solution to (\ref{eqnB}) is therefore,
\begin{eqnarray}
\lefteqn{ \frac{D (D \!-\! 3)}{(D \!-\! 1) (D \!-\! 2)} \, F_2 +
\frac{(D \!-\! 3)\, G_2}{(D \!-\! 1) (D \!-\! 2)^2} + (a a')^{D-2}
I^4[\gamma] - F_0 } \nonumber \\
& & \hspace{-.3cm} = \frac{(a a')^{D-2}}{y^{D-2}} \Biggl\{ I\Biggl[
y^{D-3} \Bigl\{ -\frac12 I^3[\beta] + (D\!+\!1) I^4[\gamma] - (D
\!-\! 1) \Phi \Bigr\} \Biggr] + K_2 \Biggr\} , \qquad
\label{finalB}
\end{eqnarray}
where the integration constant $K_2$ is chosen to prevent
$G_2(x;x')$ from having any term proportional to $1/y^{D-2}$.

We have already given the expansion (\ref{phidef}) for the de Sitter
invariant part $\Phi(y)$ of the scalar structure function.
Substituting expression (\ref{A(y)}) into (\ref{alpha}) gives the
additional expansions we require,
\begin{eqnarray}
I^3[\beta] & \!\!\!\!\!=\!\!\!\!\! & \frac{\kappa^2 H^{2D-4}
\Gamma^2(\frac{D}2)}{
(4 \pi)^D} \Biggl\{ \frac{-(\frac{4}{y})^{D-2}}{2 (D\!-\! 1)
(D \!-\! 2)} - \frac4{y} + 2 \ln\Bigl(\frac{y}{4} \Bigr) \!+\!
\dots \Biggr\}  , \qquad \label{I3beta} \\
I^4[\gamma] & \!\!\!\!\!=\!\!\!\!\! & \frac{\kappa^2 H^{2D-4}
\Gamma^2(\frac{D}2)}{
(4 \pi)^D} \Biggl\{ \frac{-D (\frac{4}{y})^{D-2}}{8 (D \!+\! 1)
(D\!-\! 1) (D \!-\! 2)} - \frac2{3 y} + \frac13 \ln\Bigl(\frac{y}{4}
\Bigr) \!+ \! \dots \Biggr\} . \qquad \label{I4gamma}
\end{eqnarray}
As before, the neglected terms have the twin properties of making
integrable contributions to the structure functions and vanishing
in $D=4$ dimensions. Substituting these expansions into expressions
(\ref{finalA}) and (\ref{finalB}) gives the following expansions
for the tensor structure functions,
\begin{eqnarray}
\lefteqn{ F_2(x;x')  = \frac{\kappa^2 (H^2 a a')^{D-2}
\Gamma^2(\frac{D}2)}{(4 \pi)^D} \Biggl\{ \frac{-(\frac{4}{y})^{D-2}}{
4 (D \!+\! 1) (D \!-\! 1) (D \!-\! 2) (D \!-\! 3)} } \nonumber \\
& & \hspace{6cm} +\! \frac23 \Bigl[\frac{1}{y} \!-\! \frac{1}{4 \!-\! y} 
\Bigr] \ln\Bigl(\frac{y}4\Bigr) \!-\! \frac13 \Psi(y) \Biggr\} , \qquad 
\label{F2exp} \\
\lefteqn{G_2(x;x') = \frac{\kappa^2 (H^2 a a')^2}{ (4\pi)^4} \Biggl\{
- 2 + \frac83 \frac{\ln(\frac{y}4)}{(4 -y)} + \frac23 \Psi(y) \Biggr\} 
. } \label{G2exp}
\end{eqnarray}
The second tensor structure function $G_2(x;x')$ is ultraviolet finite
but the first term of $F_2(x;x')$ requires the same treatment as the
most singular part of $F_0(x;x')$. The steps are the same as led to
expression (\ref{renform}) so we merely give the result,
\begin{eqnarray}
\lefteqn{ \frac{\kappa^2 (H^2 a a')^{D-2} \Gamma^2(\frac{D}2)}{(4
\pi)^D} \times -\frac1{4 (D\!+\!1) (D\!-\! 1) (D \!-\! 2) (D \!-\! 3)}
\Bigl( \frac{4}{y} \Bigr)^{D-2} } \nonumber \\
& & \hspace{-.7cm} \longrightarrow \frac{\kappa^2}{(4 \pi)^4}
\frac{\partial^2}{30} \Bigl[ \frac{\ln(\mu^2 \!\Delta x^2)}{\Delta x^2}
\Bigr] \!-\! \frac{\kappa^2 \mu^{D-4} \Gamma(\frac{D}2)}{64
\pi^{\frac{D}2}} \frac{i \delta^D(x \!-\! x')}{(D\!-\!4)
(D \!-\! 3)^2 (D \!-\! 1) (D \!+\! 1)} . \qquad \label{F2div}
\end{eqnarray}
The divergent part of expression (\ref{F2div}) can be absorbed into
the counterterm $\Delta \mathcal{L}_2 = c_2 C^{\alpha\beta\gamma\delta}
C_{\alpha\beta\gamma\delta} \sqrt{-g}$ of Ref. \cite{PW1}, see
equations (114) and (237) of that work. Hence our final renormalized
result for the first tensor structure function is,
\begin{eqnarray}
\lefteqn{ F_{2R}(x;x') = \frac{\kappa^2 (H^2 a a')^2}{(4 \pi)^4} \Biggl\{
\frac{\partial^2}{30 (H^2 a a')^2} \Bigl[ \frac{\ln(\mu^2
\Delta x^2)}{\Delta x^2} \Bigr] } \nonumber \\
& & \hspace{6cm} + \frac23 \Bigl[\frac{1}{y} \!-\! \frac{1}{4 \!-\! y} 
\Bigr] \ln\Bigl(\frac{y}4\Bigr) \!-\! \frac13 \Psi(y) \Biggr\} . 
\qquad \label{F2ren}
\end{eqnarray}
As desired, expressions (\ref{F2ren}) and (\ref{G2exp}) are
significantly simpler than the form which pertains for the manifestly
de Sitter invariant construction of Ref. \cite{PW1}.

\section{Quantum Corrections to Gravitons}\label{gravitons}

The point of this section is to re-examine the conclusion \cite{PW2}
of the de Sitter invariant analysis that the ensemble of MMC scalars
produced during inflation has no effect on dynamical gravitons at
one loop. We begin by explaining how to perturbatively formulate the
effective field equations of the Schwinger-Keldysh formalism
\cite{SK,FW}. Then we specialize to quantum correcting the mode
functions of plane wave gravitons.

\subsection{Perturbative effective field equations}

The linearized effective field equation for our graviton field
(\ref{htochi}) is,
\begin{equation}
\partial_{\alpha} \Bigl[a^2
\mathcal{L}^{\mu\nu\rho\sigma\alpha\beta}
\partial_{\beta} h_{\rho\sigma}(x) \Bigr] - \int \!\! d^4x'
\Bigl[\mbox{}^{\mu\nu} \Sigma^{\rho\sigma}\Bigr](x;x')
h_{\rho\sigma}(x') = -\frac{\kappa a^2}{2} \eta^{\mu\rho}
\eta^{\nu\sigma} T_{\rho\sigma}(x) \; . \label{lineqn}
\end{equation}
Here $T_{\rho\sigma}(x)$ is the stress-energy tensor and
$\mathcal{L}^{\mu\nu\rho\sigma\alpha\beta}$ is,
\begin{eqnarray}
\lefteqn{\mathcal{L}^{\mu\nu\rho\sigma\alpha\beta} = \frac12
\eta^{\alpha\beta} \Bigl[ \eta^{\mu (\rho} \eta^{\sigma) \nu} \!-\!
\eta^{\mu\nu} \eta^{\rho\sigma} \Bigr] } \nonumber \\
& & \hspace{3.5cm} + \frac12 \eta^{\mu\nu} \eta^{\rho (\alpha}
\eta^{\beta) \sigma} + \frac12 \eta^{\rho\sigma} \eta^{\mu (\alpha}
\eta^{\beta) \nu} - \eta^{\alpha) (\rho} \eta^{\sigma) (\mu}
\eta^{\nu) (\beta} \; . \qquad
\end{eqnarray}
With $T_{\rho\sigma} = 0 $ one can use equation (\ref{lineqn}) to
understand how inflationary particles affect the propagation of
dynamical gravitons. By setting $T_{\rho\sigma} \neq 0$ one can
study how inflation affects the force of gravity.

It is useful to re-express (\ref{lineqn}) in terms of the four
structure functions $F_0(x;x')$, $G_0(x;x')$, $F_2(x;x')$ and
$G_2(x;x')$ in our representation (\ref{ourform}) of the graviton
self-energy $[\mbox{}^{\mu\nu} \Sigma^{\rho\sigma}](x;x')$. Recall
that each term is the product of primed and unprimed, second order
differential operators acting on one of the structure functions. We
can extract the unprimed differential operator from the integration
over ${x'}^{\mu}$, and partially integrate the primed differential
operator to act on the graviton field $h_{\rho\sigma}(x')$. Carrying
this out for the $F_2$ structure function gives,
\begin{eqnarray}
\lefteqn{ \int \!\! d^4x' \, \mathcal{C}_{\alpha\beta\gamma\delta}^{
~~~~~ \mu\nu} \times {\mathcal{C}'}_{\kappa\lambda\theta\phi}^{~~~~~
\rho\sigma} \Bigl[ \eta^{\alpha\kappa} \eta^{\beta\lambda}
\eta^{\gamma\theta} \eta^{\delta\phi} iF_2(x;x') \Bigr] \times
h_{\rho\sigma}(x') }
\nonumber \\
& & \hspace{5cm} = \mathcal{C}_{\alpha\beta\gamma\delta}^{~~~~~
\mu\nu} \! \int \!\! d^4x' \, i F_2(x;x') \times
\widetilde{C}^{\alpha\beta\gamma\delta}_{\rm lin}(x') \; , \qquad
\label{F2step1} \\
& & \hspace{5cm} = -2 \partial_{\alpha} \partial_{\beta} \! \int
\!\! d^4x' \, iF_2(x;x') \times
\widetilde{C}^{\mu\alpha\nu\beta}_{\rm lin}(x') \; . \qquad
\label{F2step2}
\end{eqnarray}
Here $\kappa \widetilde{C}^{\alpha\beta\gamma\delta}_{\rm lin}(x)$
is the linearized Weyl tensor of the conformally rescaled metric
${\widetilde g}_{\mu\nu}(x) = \eta_{\mu\nu} + \kappa h_{\mu\nu}(x)$.
The transition from (\ref{F2step1}) to (\ref{F2step2}) is justified
by the tracelessness of the Weyl tensor and by its algebraic
symmetries.

With the various projectors extracted or partially integrated the
effective field equation (\ref{lineqn}) takes the form,
\begin{eqnarray}
\lefteqn{ \partial_{\alpha} \Bigl[a^2
\mathcal{L}^{\mu\nu\rho\sigma\alpha\beta} \partial_{\beta}
h_{\rho\sigma}(x) \Bigr] = -\frac{\kappa a^2}{2} \eta^{\mu\rho}
\eta^{\nu\sigma} T_{\rho\sigma}(x) } \nonumber \\
& & \hspace{.5cm} + \mathcal{F}^{\mu\nu} \!\! \int \!\! d^4x' \,
iF_0(x;x') \mathcal{R}(x') + \mathcal{G}^{\mu\nu} \!\! \int \!\!
d^4x' \, iG_0(x;x') \mathcal{S}(x') \nonumber \\
& & \hspace{1.5cm} - 2 \partial_{\alpha} \partial_{\beta} \!\! \int
\!\! d^4x' \Bigl[ iF_2(x;x') \widetilde{C}^{ \mu\alpha\nu\beta}_{\rm
lin}(x') + iG_2(x;x') \overline{ \widetilde{C}}^{~
\mu\alpha\nu\beta}_{\rm lin}\!(x') \Bigr] \nonumber \\
& & \hspace{1.6cm} + \Bigl[\eta^{\mu\nu} \partial_k \partial_{\ell}
\!-\! 2 \delta^{(\mu}_{(k} \partial^{\nu )} \partial_{\ell )} \!+\!
\delta^{(\mu}_k \delta^{\nu )}_{\ell} \partial^2 \Bigr] \!\!
\int \!\! d^4x' i G_2(x;x') \widetilde{C}^{0k0\ell}_{\rm lin}(x')
\; . \qquad \label{goodeqn}
\end{eqnarray}
We remind the reader that the spin zero projectors
$\mathcal{F}^{\mu\nu}$ and $\mathcal{G}^{\mu\nu}$ were defined in
expressions (\ref{Fansatz}) and (\ref{Gansatz}), respectively, although
they should be specialized to $D=4$ dimensions here. The quantity
$\mathcal{R}$ is $\kappa^{-1}$ times the linear part of the ($D =4$)
Ricci scalar,
\begin{equation}
\mathcal{R}(x) \equiv \partial^{\rho} \partial^{\sigma}
h_{\rho\sigma} - 6 a H \partial^{\rho} h_{0 \rho} + 12 a^2 H^2
h_{00} - \Bigl[ \partial^2 \!-\! 3 a H \partial_0 \Bigr] h \; .
\label{calR}
\end{equation}
The ``essentially spatial'' part of this is,
\begin{equation}
\mathcal{S}(x) \equiv \partial_k \partial_{\ell} h_{k\ell} - 4 a H
\partial_k h_{0k} + 6 a^2 H^2 h_{00} - \Bigl[ \nabla^2 \!-\! 2 a H
\partial_0 \Bigr] h_{kk} \; . \label{calS}
\end{equation}
The symbol $\overline{\widetilde{C}}_{\rm
lin}^{~\alpha\beta\gamma\delta}$ stands for the purely spatial
components of the linearized Weyl tensor of the conformally rescaled
metric.

Equation (\ref{goodeqn}) is general, but its use is limited because
we will never possess more than the lowest loop results for the
various structure functions. The only valid solution is to regard
$h_{\mu\nu}(x)$, and the structure functions, as the sum of results
at different loop orders,
\begin{eqnarray}
h_{\mu\nu}(x) & = & h^0_{\mu\nu}(x) + h^1_{\mu\nu}(x) +
h^2_{\mu\nu}(x) + \dots \qquad \\
F_{0,2}(x;x') & = & 0 + F^1_{0,2}(x;x') + F^2_{0,2}(x;x') + \dots
\qquad \\
G_{0,2}(x;x') & = & 0 + G^1_{0,2}(x;x') + G^2_{0,2}(x;x') + \dots
\qquad
\end{eqnarray}
Unless the stress tensor includes loop corrections from the 1PI
1-point function, we regard it as 0th order,
\begin{equation}
\partial_{\alpha} \Bigl[a^2
\mathcal{L}^{\mu\nu\rho\sigma\alpha\beta} \partial_{\beta}
h^0_{\rho\sigma}(x) \Bigr] = -\frac{\kappa a^2}{2} \eta^{\mu\rho}
\eta^{\nu\sigma} T_{\rho\sigma}(x) \; . \label{0loopeqn}
\end{equation}
The resulting zero loop field $h^0_{\mu\nu}$ then combines with the
one loop structure functions in (\ref{goodeqn}) to provide sources
for the one loop field $h^1_{\mu\nu}$,
\begin{eqnarray}
\lefteqn{\partial_{\alpha} \Bigl[a^2 \mathcal{L}^{
\mu\nu\rho\sigma\alpha\beta} \partial_{\beta} h^1_{\rho\sigma}(x)
\Bigr] } \nonumber \\
& & \hspace{.3cm} = \mathcal{F}^{\mu\nu} \!\! \int \!\! d^4x' \,
iF^1_0(x;x') \mathcal{R}^0(x') + \mathcal{G}^{\mu\nu} \!\! \int \!\!
d^4x' \, iG^1_0(x;x') \mathcal{S}^0(x') \nonumber \\
& & \hspace{1.3cm} - 2 \partial_{\alpha} \partial_{\beta} \!\! \int
\!\! d^4x' \Bigl[ iF^1_2(x;x') \widetilde{C}^{
\mu\alpha\nu\beta}_{{\rm lin}0}(x') + iG^1_2(x;x') \overline{
\widetilde{C}}^{\mu\alpha\nu\beta}_{{\rm lin}0}\!(x') \Bigr]
\nonumber \\
& & \hspace{1.8cm} + \Bigl[\eta^{\mu\nu} \partial_k \partial_{\ell}
\!-\! 2 \delta^{(\mu}_{(k} \partial^{\nu )} \partial_{\ell )} \!+\!
\delta^{(\mu}_k \delta^{\nu )}_{\ell} \partial^2 \Bigr] \!\!
\int \!\! d^4x' i G^1_2(x;x') \widetilde{C}^{0k0\ell}_{\rm lin 0}(x')
, \qquad \label{1loopeqn}
\end{eqnarray}
where $\widetilde{C}^{\alpha\beta\gamma\delta}_{{\rm lin}0}(x)$ is
the linearized conformally transformed Weyl tensor formed from the
0-th order graviton field $h^0_{\mu\nu}(x)$.

\subsection{Schwinger-Keldysh Formalism}

There are various sorts of ``effective field equations''
corresponding to different definitions of the ``effective field''
$h_{\mu\nu}(x)$. People are most familiar with the in-out effective
field equations, which describe the in-out matrix element of the
graviton field. That is indeed the best way of describing scattering
problems on a flat space background, but it has little relevance for
cosmology in which the universe began with a singularity and no one
knows how (or even if) it will end. Using the in-out effective field
equations for cosmology would have the highly undesirable effect of
making evolution at some finite time $\eta$ depend upon our
assumption about the asymptotic future. Further, because the in
state will not typically equal the out one, the in-out effective
field develops an imaginary part!

The more appropriate cosmological problem is to release the universe
in a prepared state at some finite time $\eta_i$ and then follow the
evolution of the expectation value. The Schwinger-Keldysh formalism
\cite{SK} gives the effective field equations for this situation,
and it does so in a way that is almost as simple as the Feynman
diagram technology of in-out computations. The salient features are
\cite{FW}:
\begin{itemize}
\item{The same Heisenberg field operator $\varphi(x)$ gives rise to
two dummy variables, $\phi_{\scriptscriptstyle \pm}(x)$ in the
functional integral formalism. The functional integration over
$\phi_{\scriptscriptstyle +}(x)$ implements forward time evolution
from the prepared state to some point in the future of the latest
observation, while the functional integration over
$\phi_{\scriptscriptstyle -}(x)$ implements backwards evolution to
the original state.}
\item{Each end of a Schwinger-Keldysh propagator carries a $\pm$
polarity, corresponding to which of the two dummy fields is meant.}
\item{Vertices and counterterms are either all $+$ or all $-$. The
$+$ vertices and counterterms are identical to those of the in-out
formalism, while the $-$ vertices and counterterms are conjugated.}
\item{Every 1PI N-point function of the in-out formalism gives rise
to $2^N$ Schwinger-Keldysh 1PI N-point functions, corresponding to
the two possible polarities for each leg.}
\item{It is the sum of the $++$ and $+-$ 1PI 2-point functions which
appears in linearized effective field equations such as
(\ref{lineqn}).}
\item{For the case of interest to cosmology, in which propagators
depend only on the scale factors and the de Sitter length function
(\ref{ydef}), the four Schwinger-Keldysh propagators follow from the
Feynman propagator by simply changing the $i\varepsilon$
prescription according to the rule:
\begin{eqnarray}
i\Delta_{\scriptscriptstyle ++}(x;x') & \Longrightarrow &
y_{\scriptscriptstyle ++}(x;x) \equiv H^2 a a' \Bigl[ \Vert \vec{x}
\!-\! \vec{x}' \Vert^2 - (\vert \eta \!-\! \eta' \vert \!-\! i
\varepsilon)^2 \Bigr] \; , \qquad \\
i\Delta_{\scriptscriptstyle +-}(x;x') & \Longrightarrow &
y_{\scriptscriptstyle +-}(x;x) \equiv H^2 a a' \Bigl[ \Vert \vec{x}
\!-\! \vec{x}' \Vert^2 - (\eta \!-\! \eta' \!+\! i
\varepsilon)^2 \Bigr] \; , \qquad \\
i\Delta_{\scriptscriptstyle -+}(x;x') & \Longrightarrow &
y_{\scriptscriptstyle -+}(x;x) \equiv H^2 a a' \Bigl[ \Vert \vec{x}
\!-\! \vec{x}' \Vert^2 - (\eta \!-\! \eta' \!-\! i
\varepsilon)^2 \Bigr] \; , \qquad \\
i\Delta_{\scriptscriptstyle --}(x;x') & \Longrightarrow &
y_{\scriptscriptstyle --}(x;x) \equiv H^2 a a' \Bigl[ \Vert \vec{x}
\!-\! \vec{x}' \Vert^2 - (\vert \eta \!-\! \eta' \vert \!+\! i
\varepsilon)^2 \Bigr] \; . \qquad
\end{eqnarray}}
\item{Perturbative corrections to using free vacuum as the prepared
state correspond to vertices on the initial value surface
\cite{FW,KOW}.}
\end{itemize}

It remains only to convert our previous results from the structure
functions to Schwinger-Keldysh form. At the one loop order we are
working, this is done by taking the in-out result with the
replacement $y(x;x') \rightarrow y_{\scriptscriptstyle ++}(x;x')$,
and then subtracting the in-out result with the replacement $y(x;x')
\rightarrow y_{\scriptscriptstyle +-}(x;x')$. The projectors are not
affected at all, so we merely give the three nonzero structure
functions of the Schwinger-Keldysh formalism,
\begin{eqnarray}
\lefteqn{ F^1_{0} = \frac{i \kappa^2}{576 \pi^3} \Biggl\{
\Bigl[\frac{\partial^4 \!-\! 4 H^2 a a' \partial^2}{16} \Bigr]
\Biggl[ \Bigl[ \ln\Bigl( \frac{- y}{4 a a'} \Bigr)
\!-\! 1\Bigr] \Theta \Biggr] - \frac14 H^2 a a' \ln(a a')
\partial^2 \Theta } \nonumber \\
& & \hspace{3cm} + H^4 a^2 {a'}^2 \Biggl[3 \!-\! \frac1{4 \!-\! y} 
\!+\! \frac34 (2 \!-\! y) \ln\Bigl(\frac{ -y}{4 \!-\!y} \Bigr) \Biggr] 
\Theta \Biggr\} , \qquad \label{F10} \\
\lefteqn{ F^1_{2} = \frac{i \kappa^2}{64 \pi^3} \Biggl\{ \!
\Bigl[\frac{\partial^4 \!+\! 20 H^2 a a' \partial^2}{240} \Bigr] \!
\Biggl[ \Bigl[ \ln\Bigl( \frac{- y}{4 a a'} \Bigr) - 1
\Bigr] \Theta \Biggr] \!+\! \frac{H^2 a a' \ln(a a')}{12} \, 
\partial^2 \Theta } \nonumber \\
& & \hspace{5.5cm} + H^4 a^2 {a'}^2 \Biggl[ \frac{-\frac13}{4 \!-\! y} 
\!-\! \frac16 \ln\Bigl(\frac{ -y}{4 \!-\! y} \Bigr) \Biggr] \Theta 
\Biggr\} , \qquad \label{F12} \\
\lefteqn{ G^1_{2} = \frac{i \kappa^2}{64 \pi^3} \Biggl\{H^4 a^2 {a'}^2 
\Biggl[\frac{\frac43}{4 \!-\! y} \!+\! \frac13 \ln\Bigl(\frac{ -y}{4 
\!-\! y} \Bigr) \Biggr] \Theta \Biggr\} . } \label{G12}
\end{eqnarray}
In these expressions the symbol $\Theta$ stands for the
$\theta$-function which enforces causality,
\begin{equation}
\Theta \equiv \theta\Bigl( \Delta \eta - \Vert \vec{x} \!-\!
\vec{x}' \Vert\Bigr) \qquad , \qquad \Delta \eta \equiv \eta \!-\!
\eta' \; , \label{Theta}
\end{equation}
and we remind the reader that $-y(x;x')$ is,
\begin{equation}
-y(x;x') = H^2 a a' \Bigl[ \Delta \eta^2 - \Vert \vec{x} \!-\!
\vec{x}' \Vert^2 \Bigr] \; .
\end{equation}
Note also that we have set our renormalization scale to $\mu =
\frac12 H$ in order to facilitate comparison with the results of
Refs. \cite{PW1,PW2}.

Several points about expressions (\ref{F10}-\ref{G12}) deserve
comment. First, the fact that each of the Schwinger-Keldysh structure
functions is pure imaginary means that the effective field equation
(\ref{1loopeqn}) is pure real. This is an important feature of the
Schwinger-Keldysh effective field equations which is not shared by
the more familiar, in-out effective field equations. A similarly
distinctive feature is the causality enforcing $\theta$-function
(\ref{Theta}). One consequence of this causality is that partially
integrating spatial derivatives can produce no surface terms provided
the interaction begins at some finite time. Partial integration of
time derivatives produces no surface terms in the future but it can
give rise to surface terms at the initial time. Because perturbative
corrections to the initial state also produce initial time surface
integrals it is usual to assume that they cancel the surface terms
produced by desired partial time integrations \cite{KOW}, such as
those involved in reflecting the primed projectors off of the
structure functions and onto the graviton field. However, this has
not been checked.

\subsection{Source for dynamical gravitons}

To study dynamical gravitons we set the stress tensor to zero in
equation (\ref{0loopeqn}). The general solution can be expressed as
a superposition of transverse-traceless, spatial plane wave
gravitons of the form,
\begin{equation}
h^0_{\mu\nu}(x) = \epsilon_{\mu\nu}(\vec{k}) u_0(\eta,k) e^{i
\vec{k} \cdot \vec{x}} \quad , \quad u_0(\eta,k) = \frac{H}{\sqrt{2
k^3}} \Bigl[ 1 - \frac{ik}{H a} \Bigr] e^{-ik \eta} \; .
\label{0sol}
\end{equation}
The polarization tensor $\epsilon_{\mu\nu}(\vec{k})$ is identical to
the one usually employed in flat space. In particular, its temporal
components vanish, and it is transverse and traceless,
\begin{equation}
\epsilon_{\mu 0}(\vec{k}) = 0 \qquad , \qquad k_i
\epsilon_{ij}(\vec{k}) = 0 \qquad , \qquad \epsilon_{ii}(\vec{k}) =
0 \; . \label{eps}
\end{equation}
Taken with expressions (\ref{calR}-\ref{calS}), these facts
demonstrate the vanishing of the spin zero contributions to one loop
effective field equation (\ref{1loopeqn}),
\begin{equation}
h^0_{\mu\nu}(x) = \epsilon_{\mu\nu}(\vec{k}) u_0(\eta,k) e^{i
\vec{k} \cdot \vec{x}} \qquad \Longrightarrow \qquad
\mathcal{R}^0(x) = 0 = \mathcal{S}^0(x) \; .
\end{equation}

For the zeroth order solution (\ref{0sol}-\ref{eps}) the only
nonzero components of the linearized Weyl tensor are (up to index
permutations),
\begin{eqnarray}
\widetilde{C}^{0i0j}_{{\rm lin}0}(x) & = & -\frac14 \epsilon^{ij}
\times (\partial_0^2 \!-\! k^2) u_0(\eta,k) e^{i \vec{k} \cdot
\vec{x}} \; , \qquad \label{C1} \\
\widetilde{C}^{0ijk}_{{\rm lin}0}(x) & = & -\frac{i}2
\Bigl(\epsilon^{ij} k^k \!-\! \epsilon^{ik} k^j \Bigr) \times
\partial_0 u_0(\eta,k) e^{i \vec{k} \cdot \vec{x}} \; , \qquad
\label{C2} \\
\widetilde{C}^{ijk\ell}_{{\rm lin}0}(x) & = & \frac12 \Bigl(
\epsilon^{ik} k^j k^{\ell} \!-\! \epsilon^{kj} k^{\ell} k^i \!+\!
\epsilon^{j\ell} k^i k^k \!-\! \epsilon^{\ell i} k^k k^j \Bigr)
\times u_0(\eta,k) e^{i \vec{k} \cdot \vec{x}} \nonumber \\
& & \hspace{-.5cm} -\frac14 \Bigl(\epsilon^{ik} \delta^{j\ell}
\!-\! \epsilon^{kj} \delta^{\ell i} \!+\! \epsilon^{j\ell}
\delta^{i k} \!-\! \epsilon^{\ell i} \delta^{k j} \Bigr)\times
(\partial_0^2 \!+\! k^2) u_0(\eta,k) e^{i \vec{k} \cdot \vec{x}}
\; . \qquad \label{C3}
\end{eqnarray}
Now make a $3 + 1$ decomposition of the first spin two contribution
to (\ref{1loopeqn}),
\begin{eqnarray}
\lefteqn{2 \partial_{\alpha} \partial_{\beta} \!\! \int \!\! d^4x'
iF_2^1(x;x') \widetilde{C}^{\alpha\mu\beta\nu}_{{\rm lin}0}(x') = 2
\partial_0^2 \!\! \int \!\! d^4x' \, i F_2^1(x;x') \widetilde{C}^{0\mu
0\nu}_{{\rm lin}0}(x') } \nonumber \\
& & \hspace{-.3cm} - 4 \partial_0 \partial_k \!\! \int \!\! d^4x' \,
i F_2^1(x;x') \widetilde{C}^{0 (\mu \nu) k}_{{\rm lin}0}(x') + 2
\partial_k \partial_{\ell} \!\! \int \!\! d^4x' \, i F_2^1(x;x')
\widetilde{C}^{k \mu \ell \nu}_{{\rm lin}0}(x') \; , \qquad \\
& & \hspace{-.5cm} = 2 \partial_0^2 \!\! \int \!\! d^4x' \, i
F_2^1(x;x') \widetilde{C}^{0\mu 0\nu}_{{\rm lin}0}(x') \nonumber \\
& & \hspace{-.3cm} - 4i \partial_0 k_k \!\! \int \!\! d^4x' \,
i F_2^1(x;x') \widetilde{C}^{0 (\mu \nu) k}_{{\rm lin}0}(x') - 2 k_k
k_{\ell} \!\! \int \!\! d^4x' \, i F_2^1(x;x') \widetilde{C}^{k \mu
\ell \nu}_{{\rm lin}0}(x') \; . \qquad \label{3+1}
\end{eqnarray}
The analogous expansion for the second spin two contribution is just
the last term. From relations (\ref{C1}-\ref{C3}) we see that only
the third of the $\widetilde{C}^{0k0\ell}_{\rm lin 0}$ terms survives,
\begin{eqnarray}
\lefteqn{ \Bigl[\eta^{\mu\nu} \partial_k \partial_{\ell}
\!-\! 2 \delta^{(\mu}_{(k} \partial^{\nu )} \partial_{\ell )} \!+\!
\delta^{(\mu}_k \delta^{\nu )}_{\ell} \partial^2 \Bigr] \!\! \int
\!\! d^4x' i G^1_2(x;x') \widetilde{C}^{0k0\ell}_{\rm lin}(x') }
\nonumber \\
& & \hspace{5cm} = -(\partial_0^2 + k^2) \!\! \int \!\! d^4x'
i G^1_2(x;x') \widetilde{C}^{0\mu0\nu}_{\rm lin}(x') \; . \qquad
\end{eqnarray}
There is no contribution when either of the indices $\mu$ or $\nu$
is temporal. When $\mu = i$ and $\nu = j$ equation (\ref{1loopeqn})
reads,
\begin{eqnarray}
\lefteqn{\partial_{\alpha} \Bigl[a^2 \mathcal{L}^{ij
\rho\sigma\alpha\beta} \partial_{\beta} h^1_{\rho\sigma}(x) \Bigr] =
2 \epsilon^{ij} k^2 \partial_0 \!\! \int \!\! d^4x' \, i F_2^1(x;x')
\partial'_0 u_0(\eta',k) e^{i \vec{k} \cdot \vec{x}'} } \nonumber \\
& & \hspace{.3cm} + \frac12 \epsilon^{ij} (\partial_0^2 \!-\! k^2)
\!\! \int \!\! d^4x' \Bigl[iF_2^1(x;x') \!+\! \frac{i}2 G^1_2(x;x')
\Bigr] ({\partial_0'}^2 \!-\! k^2) u_0(\eta',k) e^{i \vec{k} \cdot
\vec{x}'} \! . \qquad \label{1loopT}
\end{eqnarray}

In view of equation (\ref{1loopT}) we may assume that the one loop
correction to the graviton field takes the same form as
(\ref{0sol}),
\begin{equation}
h^1_{\mu\nu}(x) = \epsilon_{\mu\nu}(\vec{k}) u_1(\eta,k) e^{i
\vec{k} \cdot \vec{x}} \; . \label{1sol}
\end{equation}
(This form obviously pertains to all orders.) Substituting
(\ref{1sol}) into (\ref{1loopT}), and factoring out the polarization
tensor and the trivial spatial dependence, allows us to read off an
equation for the one loop correction to mode function,
\begin{eqnarray}
\lefteqn{ -\frac12 a^2 \Bigl[ \partial_0^2 \!+\! 2 a H \partial_0
\!+\! k^2 \Bigr] u_1(\eta,k) = 2 k^2 \partial_0 \!\! \int \!\! d^4x' \,
iF^1_2(x;x') \partial'_0 u_0(\eta',k) e^{-i \vec{k} \cdot \Delta
\vec{x}} } \nonumber \\
& & \hspace{.2cm} +\frac12 (\partial_0^2 \!-\! k^2 ) \!\! \int \!\!
d^4x' \Bigl[ iF_2^1(x;x') \!+\! \frac{i}2 G^1_2(x;x') \Bigr]
({\partial'_0}^2 \!-\! k^2) u_0(\eta',k) e^{-i \vec{k} \cdot \Delta
\vec{x}} , \qquad \label{modeqn}
\end{eqnarray}
where $\Delta \vec{x} \equiv \vec{x} - \vec{x}'$. The 0th order
solution (\ref{0sol}) implies an important simplification,
\begin{equation}
(\partial_0^2 - k^2) u_0(\eta,k) = -2 i k \times \partial_0
u_0(\eta,k) = -2 i k \times \frac{H}{\sqrt{2 k^3}} \Bigl[
\frac{-k^2}{H a} \Bigr] e^{-i k \eta} \; . \label{uident}
\end{equation}
Relation (\ref{uident}) allows us to combine the various $F^1_2$
terms in (\ref{modeqn}),
\begin{eqnarray}
\lefteqn{ -\frac12 a^2 \Bigl[ \partial_0^2 \!+\! 2 a H \partial_0
\!+\! k^2 \Bigr] u_1(\eta,k) = k (\partial_0 \!+\! i k)^2
\!\!\! \int \!\! d^4x' F^1_2(x;x') \partial'_0 u_0(\eta',k)
e^{-i \vec{k} \cdot \Delta \vec{x}} } \nonumber \\
& & \hspace{4cm} + \frac12 k (\partial_0^2 \!-\! k^2) \!\! \int
\!\!\! d^4x' G_2^1(x;x') \partial'_0 u_0(\eta',k) e^{-i \vec{k}
\cdot \Delta \vec{x}} . \qquad \label{newmodeqn}
\end{eqnarray}
Note that we have not yet made any assumption --- beyond
homogeneity and isotropy --- about the form of the tensor
structure functions, so equations (\ref{modeqn}) and
(\ref{newmodeqn}) are correct for any contribution to the
graviton self-energy, including that of gravitons.

Because we have not corrected the initial state equation \cite{KOW},
(\ref{modeqn}) and (\ref{newmodeqn}) can only be reliably used to
infer secular growth, if there is any. Based on the experience of
\cite{PMTW}, the minimum interesting secular dependence would be
$u_1(\eta,k) \sim \ln(a)/a^2$, which results in no correction to
the tensor power spectrum but does cause the electric components of
the one loop linearized Weyl tensor to grow like $\kappa^2 H^2
\ln(a)$ relative to the classical result (\ref{C1}). To get
$u_1(\eta,k) \sim \ln(a)/a^2$ would require the source terms on the
right hand side of (\ref{modeqn}) and (\ref{newmodeqn}) to grow
like $a^2 \ln(a)$.

The simplest part of the structure functions to analyze is the one
implicit in the flat space limit \cite{LW2},
\begin{equation}
\Bigl( F_2^1(x;x') \Bigr)_{\rm flat} = \frac{i \kappa^2 \partial^4}{2^{10}
\!\cdot\! 3 \!\cdot\! 5 \!\cdot\! \pi^3} \Biggl\{ \theta(\Delta \eta
\!-\! \Delta x) \Biggl[ \ln\Bigl[H^2 (\Delta \eta^2 \!-\! \Delta x^2)
\Bigr] \!-\! 1 \Biggr] \Biggr\} . \label{F2flat}
\end{equation}
Substituting this in (\ref{newmodeqn}) and performing the integrations
gives,
\begin{eqnarray}
\lefteqn{k (\partial_0 \!+\! i k)^2 \!\!\! \int \!\! d^4x'
\Bigl(F^1_2(x;x') \Bigr)_{\rm flat} \partial'_0 u_0(\eta',k)
e^{-i \vec{k} \cdot \Delta \vec{x}} } \nonumber \\
& & \hspace{1.5cm} = -\frac{\kappa^2 (\partial_0^2 \!+\! k^2)^2}{2^8
\!\cdot\! 3 \!\cdot\! 5 \!\cdot\! \pi^2} \Biggl\{ \Biggl[ 2\ln(H
\Delta \eta_i) + \int_0^{2k \Delta \eta_i} \!\!\!\!\!\! dt \Bigl[
\frac{e^{it} \!-\!1}{t}\Bigr] \Biggr] u(\eta,k) \Biggr\} , \qquad
\label{Sflat}
\end{eqnarray}
where $H \Delta \eta_i = 1 -\frac1{a}$. Expression (\ref{Sflat})
approaches a constant at late times, which would induce irrelevant
corrections of the form $u_1(\eta,k) \sim \frac1{a^4}$. However,
expression (\ref{Sflat}) can probably be completely cancelled by the
same state correction which eliminates its flat space cousin \cite{LW2}.

We might term the remaining parts of the structure functions as ``dS''
because they contain one or two multiplicative factors of the de Sitter
Hubble parameter and scale factors, $H^2 a a'$. After a number of
tedious manipulations these terms can be expressed as the sum of three
double integrals,
\begin{eqnarray}
\lefteqn{k (\partial_0 \!+\! i k)^2 \!\!\! \int \!\! d^4x'
\Bigl[F^1_2(x;x') + \frac12 G^1_2(x;x') \Bigr]_{\rm dS} \partial'_0 
u_0(\eta',k) e^{-i \vec{k} \cdot \Delta \vec{x}} } \nonumber \\
& & \hspace{5.5cm} - i k^2 \partial_0 \!\! \int \!\!\! d^4x' \,
G_2^1(x;x') \partial'_0 u_0(\eta',k) e^{-i \vec{k} \cdot \Delta 
\vec{x}} \nonumber \\
& & \hspace{-.5cm} = \frac{i\kappa^2 H k^2 u_0(0,k)}{96 \pi^2} \, 
(\partial_0 \!+\! i k)^2 a \!\! \int_{\eta_i}^{\eta} \!\!\!\! d\eta' 
e^{-ik \eta'} \Biggl\{\sin(k \Delta \eta) \ln\Bigl( \frac{\Delta 
\eta^2}{\eta \eta'}\Bigr) \nonumber \\
& & \hspace{1.5cm} + \sin\Bigl[ k ( \eta \!+\! \eta')\Bigr] 
\ln\Bigl(\frac{\eta'}{\eta}\Bigr) + \!\! \int_0^1 \!\! \frac{dt}{t} 
\Biggl[\sin\Bigl[k \Delta \eta (1 \!-\! 2 t)\Bigr] \!-\! 
\sin(k \Delta \eta) \nonumber \\
& & \hspace{3cm} - \sin\Bigl[ k(\eta \!+\! \eta' \!-\! 2 \eta t)
\Bigr] \!+\! \sin\Bigl[ k (\eta \!+\! \eta' \!-\! 2 \eta' t)\Bigr] 
\Biggr] \Biggr\} \nonumber \\
& & \hspace{-.3cm} + \frac{\kappa^2 H k^3 u_0(0,k)}{24 \pi^2} \, 
\partial_0 \, a \!\! \int_{\eta_i}^{\eta} \!\!\!\! d\eta' 
e^{-ik \eta'} \Biggl\{\Biggl[ \sin\Bigl[k ( \eta \!+\! \eta')\Bigr]  
\!-\! \frac{\sin(k \eta)}{k \eta} \sin(k \eta') \Biggr] 
\ln\Bigl(\frac{\eta'}{\eta}\Bigr) \nonumber \\
& & \hspace{3cm} + \!\! \int_0^1 \!\! \frac{dt}{t} \Biggl[ 
\sin\Bigl[ k (\eta \!+\! \eta' \!-\! 2 \eta' t)\Bigr] - \sin\Bigl[ k
(\eta \!+\! \eta' \!-\! 2 \eta t)\Bigr] \Biggr] \Biggr\} \nonumber \\
& & \hspace{-.3cm} + \frac{\kappa^2 H^2 k u_0(0,k)}{48 \pi^2} \, 
\partial_0 \, a^2 \!\! \int_{\eta_i}^{\eta} \!\!\!\! d\eta' 
\frac{e^{-ik \eta'}}{\eta'} \Biggl\{T(k \Delta \eta) 
\ln\Bigl(\frac{\Delta \eta^2}{{\eta'}^2} \Bigr) \nonumber \\
& & \hspace{1cm} + T(k \eta) \cos(k \eta') 
\ln\Bigl( \frac{{\eta'}^2}{\eta^2}\Bigr) + 
\!\! \int_0^1 \!\! \frac{dt}{t} \Biggl[ T\Bigl[
k \Delta \eta (1 \!-\! 2 t)\Bigr] \!-\! T(k \Delta \eta) \nonumber \\
& & \hspace{4cm} - T\Bigl[ k (\eta \!+\! \eta' \!-\! 2 \eta t)\Bigr] 
+ T\Bigl[k (\eta \!+\! \eta' \!-\! 2 \eta' t)\Bigr] \Biggr] \Biggr\} 
, \qquad \label{dS}
\end{eqnarray}
where $T(x) \equiv \sin(x) - x \cos(x)$. Further simplifications are
possible but the small $\eta$ behavior of each integrand already 
shows that none of the three terms can grow faster than $a^2$ after
the derivatives have been acted. Hence we confirm the conclusion of 
\cite{PW2} that the inflationary production of massless, minimally 
coupled scalars makes no significant corrections to dynamical 
gravitons at one loop order. Another point to note is that acting 
the two derivatives on the logarithm of $\Delta \eta^2$ in the first 
integral produces terms which diverge at the initial time, which is 
an indication that they should be absorbed into perturbative 
corrections to the initial state \label{KOW}. 

\section{Discussion}\label{discuss}

We have developed a noncovariant, but simple, representation for the
tensor structure of matter contributions to the self-energy of a
conformally re-scaled graviton, $g^{\rm full}_{\mu\nu} \equiv a^2
(\eta_{\mu\nu} + \kappa h_{\mu\nu})$, where $a = -1/H\eta$ is the de
Sitter scale factor and $\kappa^2 \equiv 16\pi G$ is the loop-counting
parameter of quantum gravity. Our representation is a sum of four terms,
each of which consists of a transverse, 4th order differential operator
acting on a structure function,
\begin{eqnarray}
\lefteqn{-i\Bigl[ \mbox{}^{\mu\nu} \Sigma^{\rho\sigma}\Bigr](x;x') =
\mathcal{F}^{\mu\nu}(x) \times \mathcal{F}^{\rho\sigma}(x') \Bigl[
F_0(x;x') \Bigr] } \nonumber \\
& & \hspace{-.5cm} + \mathcal{G}^{\mu\nu}(x) \times
\mathcal{G}^{\rho\sigma}(x') \Bigl[ G_0(x;x') \Bigr] +
\mathcal{F}^{\mu\nu\rho\sigma} \Bigl[ F_2(x;x') \Bigr] +
\mathcal{G}^{\mu\nu\rho\sigma} \Bigl[ G_2(x;x') \Bigr] \; . \qquad
\label{oursigma}
\end{eqnarray}
There are two scalar contributions and two tensor ones. The scalar operator
$\mathcal{F}^{\mu\nu}$ is given in expression (\ref{Fansatz}), while
$\mathcal{G}^{\mu\nu}$ is given in (\ref{Gansatz}). The transverse and
traceless operators $\mathcal{F}^{\mu\nu\rho\sigma}$ and
$\mathcal{G}^{\mu\nu\rho\sigma}$ --- see expressions (\ref{calF}) and
(\ref{calG}), respectively --- are constructed by contracting different
8-index tensors into the product of primed and unprimed Weyl operators
$\mathcal{C}_{\alpha\beta\gamma\delta}^{~~~~~ \mu\nu}$, for which see
(\ref{calC}). The two $F$-type structure functions, $F_0(x;x')$ and
$F_2(x;x')$ survive in the flat space limit, whereas $G_0(x;x')$ and
$G_2(x;x')$ vanish in that limit.

In section \ref{general} we have also given a general procedure for
determining the four structure functions from the primitive form of the
self-energy. One first finds the scalar structure functions $F_0(x;x')$
and $G_0(x;x')$ by taking the trace on one index group --- say $\rho$
and $\sigma$ --- and then examining the $\mu\nu = 0i$ and $\mu\nu = jk$
components. The resulting expressions for $F_0(x;x')$ and $G_0(x;x')$
are (\ref{F0sol}) and (\ref{G0sol}), respectively. Once the scalar
structure functions are known one finds the tensor structure functions
by examining the case for which all indices differ. The resulting
expressions for $F_2(x;x')$ and $G_2(x;x')$ are (\ref{eqn3}) and
(\ref{eqn4}).

As an example, we constructed the structure functions for the one loop
contribution from a massless, minimally coupled scalar \cite{PW1}. For
this case the structure function $G_0(x;x')$ happens to vanish. Our
results for the other structure functions are expression (\ref{F0R})
for $F_0(x;x')$, expression (\ref{F2ren}) for $F_2(x;x')$, and
expression (\ref{G2exp}) for $G_2(x;x')$. Because this model shows no
physical breaking of de Sitter invariance one could have employed a de
Sitter covariant representation of the self-energy, however, our
noncovariant representation is much simpler. Any doubts about this can
be quickly settled by comparison with the {\it full page} expressions
(234) and (258) for the de Sitter invariant structure functions
\cite{PW1}.

Our noncovariant representation is also much easier to use in the
effective field equations than the covariant one. Expression
(\ref{1loopeqn}) gives a formula for the one loop graviton field.
When specialized to the case of dynamical gravitons it becomes
(\ref{newmodeqn}). One nice property of this representation is
that surface terms really fall off like powers of the inflationary
scale factor, whereas this is not always true in de Sitter invariant
formulations \cite{LPW1}. We used this property to check that
massless, minimally coupled scalars really have no significant
effect on dynamical gravitons at one loop. The previous de Sitter
invariant analysis was based on the assumption that a certain
surface term can be ignored, either because it falls off or because
it can be absorbed into corrections of the initial state \cite{PW2}.
Our new result confirms this assumption.

Although the point of our paper was to develop a new representation
for the graviton self-energy, it is worth commenting on the physics
we found in applying it. We used the new formalism to demonstrate the
absence of any effect on dynamical gravitons from massless, minimally
coupled (MMC) scalars. This is not because there are no scalars. 
Inflation produces a veritable sea of infrared scalars which mediate 
significant effects on themselves \cite{phi4}, on photons 
\cite{SQED,PP}, and on fermions \cite{Yukawa}. The difference between 
those cases and the null result we found for gravitons is the presence 
of non-derivative interactions. The inflationary production of MMC 
scalars engenders a steady increase in the magnitude of the scalar 
field strength, so particles which couple to undifferentiated scalars 
experience an effect. In contrast, gravitons couple minimally only to 
{\it differentiated} scalars, and there is no build-up of that.

Although we have not done so, one can also employ the graviton 
self-energy to study the force of gravity. In this case there must be
some effect from MMC scalars because even the flat space limit shows 
a correction to the Newtonian potential at one loop \cite{flatscal},
\begin{equation}
\Phi(r) = \frac{GM}{r} \Bigl\{ 1 + \frac{\hbar G}{20 \pi c^3 r^2} +
\dots \Bigr\} \; . \label{flatphi}
\end{equation}
The physical origin of this effect is that the classical potential of 
a point mass tends to attract virtual scalars, which adds to the 
source. During inflation there should be an additional, secular 
effect as newly created scalars accrete onto the source. It is 
conceivable that this secular effect shows up as a one loop 
correction of the form,
\begin{equation}
\Delta \Phi(t,r) = \frac{GM}{r} \times \Bigl({\rm Constant}\Bigr)
\times \frac{\hbar G H^2}{c^5} \times \ln(a) \; . \label{dSphi}
\end{equation}
At this order such a correction would be indistinguishable from a 
secular increase of the Newton constant, which would be a fascinating
result. We now have a formalism of sufficient power and simplicity to
confirm or refute this possibility. We can also identify what part of
the structure functions is most likely to cause it: the terms $6 
\ln(\frac{y}4)$ and $-\frac32 y \Psi(y)$ on the second line of 
expression (\ref{F0R}) for the scalar structure function $F_0(x;x')$. 
Recall that $F_0(x;x')$ drops out of the effective field equations 
for dynamical gravitons, but it contributes to the equations for the 
force of gravity. Had either of the tensor structure functions 
(\ref{G2exp}) and (\ref{F2ren}) possessed the same sort of terms we 
would have found secular enhancements in the curvature carried by 
dynamical gravitons.  

Finally, two extensions of our formalism are worth noting. First, this 
same representation (\ref{oursigma}) can be used as well for graviton
contributions in transverse gauge. In other gauges --- for example,
\cite{TW2} --- the self-energy is not transverse and one would need
to revise the scalar terms but the form of the spin two contributions
would not be changed. Second, the same representation (\ref{oursigma})
should apply to any homogeneous, iso\-trop\-ic and spatially flat
geometry, with only some slight generalizations to the two scalar
terms.

\vskip 1cm

\centerline{\bf Acknowledgements}

It is a pleasure to acknowledge conversations on this subject with
P. J. Mora. This work was partially supported by the Eberly Research
Funds at Pennsylvania State University, by the D-ITP consortium, a
program of the NWO that is funded by the Dutch Ministry of Education,
Culture and Science (OCW), by NSF grant PHY-1205591, and by the
Institute for Fundamental Theory at the University of Florida. KEL
and RPW are grateful to the University of Utrecht for its hospitality
during the inception of this project.

\section{Appendix: The Projectors $\mathcal{F}^{\mu\nu\rho\sigma}$
and $\mathcal{G}^{\mu\nu\rho\sigma}$}\label{appendix}

Substituting expressions (\ref{calC}-\ref{calD2}) into (\ref{calF})
and performing some tedious algebra gives a relatively simple form
for $\mathcal{F}^{\mu\nu\rho\sigma}$ that is manifestly transverse
on each index group,
\begin{equation}
\mathcal{F}^{\mu\nu\rho\sigma} = \frac12 \Bigl(
\mathcal{P}^{\mu\rho} \mathcal{P}^{\nu\sigma} \!+\!
\mathcal{P}^{\mu\sigma} \mathcal{P}^{\nu\rho} \Bigr) -
\frac{4}{D\!-\!2} \mathcal{D}_{\alpha\beta}^{~~~ \mu\nu}
{\mathcal{D}'}^{\alpha\beta\rho\sigma} + \frac{2
\mathcal{D}^{\mu\nu} {\mathcal{D}'}^{\rho\sigma}}{(D\!-\!1)
(D\!-\!2)} \; . \label{Fexp0}
\end{equation}
Here the projector $\mathcal{P}^{\mu\rho}$ is the same one that acts
on the structure function $F(x;x')$ in the vacuum polarization,
\begin{equation}
\mathcal{P}^{\mu\rho} \equiv \eta^{\mu\rho} \partial' \!\cdot\!
\partial - {\partial'}^{\mu} \partial^{\rho} \; . \label{Fexp1}
\end{equation}
Expanding out the first term on the right hand side of (\ref{Fexp0})
gives,
\begin{equation}
\frac12 \Bigl( \mathcal{P}^{\mu\rho} {\mathcal{P}'}^{\nu\sigma}
\!+\! \mathcal{P}^{\mu\sigma} {\mathcal{P}'}^{\nu\rho} \Bigr) =
\eta^{\mu(\rho} \eta^{\sigma)\nu} (\partial \cdot \partial^\prime)^2
- 2 \partial^{\prime(\mu} \eta^{\nu)(\rho} \partial^{\sigma)}
\partial \cdot \partial^\prime + \partial^{\prime\mu}
\partial^{\prime\nu} \partial^\rho \partial^\sigma \; .
\label{Fexp2}
\end{equation}
A similarly explicit form for the contraction in the middle term of
(\ref{Fexp0}) is,
\begin{eqnarray}
\lefteqn{ \hspace{-.3cm} 4 \mathcal{D}_{\alpha\beta}^{~~~ \mu\nu}
{\mathcal{D}}'^{\alpha\beta\rho\sigma} = \eta^{\mu\nu} \eta^{\rho\sigma}
(\partial \!\cdot\! \partial^\prime)^2 \!+\! \eta^{\rho(\mu}
\eta^{\nu)\sigma} \partial^2 {\partial^\prime}^2 + \eta^{\mu\nu}
\Bigl[\partial^\rho \partial^{\sigma} {\partial^\prime}^2 \!-\! 2
\partial^{(\rho} \partial^{\prime \sigma)} \partial \!\cdot\!
\partial' \Bigr] } \nonumber \\
& & + \eta^{\rho\sigma} \Bigl[ \partial^{\prime\mu}
\partial^{\prime\nu} \partial^2 \!-\! 2 \partial^{(\mu}
\partial^{\prime \nu)} \partial \!\cdot\! \partial' \Bigr]
- 2 \partial^{(\mu} \eta^{\nu)(\rho} \partial^{\sigma)}
{\partial^\prime}^2 - 2 \partial^{\prime(\mu} \eta^{\nu)(\rho}
\partial^{\prime\sigma)} \partial^2 \nonumber \\
& & \hspace{5cm} + 2 \partial^{(\mu} \eta^{\nu)(\rho}
\partial^{\prime\sigma)} \partial \!\cdot\!
\partial^\prime + 2 \partial^{\prime(\mu} \partial^{\nu)}
\partial^{(\rho} \partial^{\prime\sigma)} \; . \qquad \label{Fexp3}
\end{eqnarray}

The ``essentially spatial'' projector
$\mathcal{G}^{\mu\nu\rho\sigma}$ requires contractions of the
linearized Riemann operator (\ref{calD6}) into one or two spatial
metrics $\overline{\eta}^{\mu\nu} \equiv \eta^{\mu\nu}
+\delta^{\mu}_0 \delta^{\nu}_0$,
\begin{eqnarray}
\mathcal{D}_{1 \beta\delta}^{~~~~ \mu\nu} & \equiv &
\eta^{\alpha\gamma} \mathcal{D}_{\overline{\alpha \beta \gamma
\delta}}^{~~~~~ \mu\nu} = -\frac12 \Bigl[ \eta^{\mu\nu}
\overline{\partial}_{\beta} \overline{\partial}_{\delta} \!-\! 2
\partial^{(\mu} {\overline{\delta}}^{\nu)}_{~ (\beta}
{\overline{\partial}}_{\delta )} \!+\!
{\overline{\delta}}^{(\mu}_{~\beta} {\overline{\delta}}^{\nu)}_{~
\delta} \partial^2\Bigr] \; , \qquad \\
\mathcal{D}_{1}^{~ \mu\nu} & \equiv & \eta^{\alpha\gamma}
\overline{\eta}^{\beta\delta} \mathcal{D}_{\alpha \beta \gamma
\delta}^{~~~~~ \mu\nu} = -\frac12 \Bigl[ \eta^{\mu\nu} \nabla^2
\!-\! 2 \partial^{(\mu} {\overline{\partial}}^{\nu)} \!+\!
{\overline{\eta}}^{\mu\nu} \partial^2\Bigr] \; , \qquad \\
\mathcal{D}_{2}^{~ \mu\nu} & \equiv & \overline{\eta}^{\alpha\gamma}
\overline{\eta}^{\beta\delta} \mathcal{D}_{\alpha \beta \gamma
\delta}^{~~~~~ \mu\nu} = \overline{\partial}^{\mu}
\overline{\partial}^{\nu} \!-\! \overline{\eta}^{\mu\nu} \nabla^2 \;
. \qquad
\end{eqnarray}
Substituting (\ref{calC}-\ref{calD2}) into (\ref{calG}) and working
through the tensor algebra gives,
\begin{eqnarray}
\lefteqn{ \mathcal{G}^{\mu\nu\rho\sigma} = \frac12 \Bigl[
\mathcal{P}_1^{~ \mu\rho} \mathcal{P}_1^{~ \nu\sigma} \!+\!
\mathcal{P}_1^{~ \mu\sigma} \mathcal{P}_1^{~ \nu\rho} \Bigr] +
\frac{4(D \!-\! 3)}{(D \!-\! 2)^2} \,
\mathcal{D}_{\overline{\alpha\beta}}^{~~~ \mu\nu}
{\mathcal{D}'}^{\alpha\beta\rho\sigma} } \nonumber \\
& & \hspace{0cm} -\frac4{D \!-\! 2} \Bigl[ \mathcal{D}_{1
\alpha\beta}^{~~~ ~\mu\nu} {\mathcal{D}'}^{\alpha\beta\rho\sigma}
\!+\! \mathcal{D}_{\alpha\beta}^{~~ \mu\nu} {\mathcal{D}'}_1^{~
\alpha\beta\rho\sigma} \Bigr] + \frac4{(D \!-\! 2)^2} \,
\mathcal{D}_1^{~ \mu\nu} {\mathcal{D}'}_1^{~\rho\sigma}
\nonumber \\
& & \hspace{.5cm} + \frac{[ 2 \mathcal{D}^{\mu\nu}
{\mathcal{D}'}^{\rho\sigma} \!-\! 4 \mathcal{D}_1^{~ \mu\nu}
{\mathcal{D}'}^{\rho\sigma} \!-\! 4 \mathcal{D}^{\mu\nu}
{\mathcal{D}'}_1^{~ \rho\sigma} \!+\! 2 \mathcal{D}^{\mu\nu}
{\mathcal{D}'}_2^{~ \rho\sigma} \!+\! 2 \mathcal{D}_2^{~ \mu\nu}
{\mathcal{D}'}^{\rho\sigma}]}{(D \!-\! 1) (D \!-\! 2)} \; , \qquad
\label{Gexp0}
\end{eqnarray}
where $\mathcal{P}_1^{~ \mu\rho}$ is the same spatial transverse
projector that appears on the structure function $G(x;x')$ in the
vacuum polarization,
\begin{equation}
\mathcal{P}_1^{~\mu\rho} \equiv \overline{\eta}^{\mu\rho} \nabla'
\!\cdot\! \nabla - {\overline{\partial}'}^{\mu}
\overline{\partial}^{\rho} \; . \label{Gexp1}
\end{equation}
The explicit form of the first term on the right of (\ref{Gexp0})
is,
\begin{equation}
\frac12 \Bigl( \mathcal{P}_1^{~\mu\rho}
{\mathcal{P}'}_1^{~\nu\sigma} \!+\! \mathcal{P}_1^{~\mu\sigma}
{\mathcal{P}'}_1^{~\nu\rho} \Bigr) = \overline{\eta}^{\mu(\rho}
\overline{\eta}^{\sigma)\nu} (\nabla \cdot \nabla^\prime)^2 - 2
\overline{\partial}^{\prime(\mu} \overline{\eta}^{\nu)(\rho}
\overline{\partial}^{\sigma)} \nabla \cdot \nabla^\prime +
\overline{\partial}^{\prime\mu} \overline{\partial}^{\prime\nu}
\overline{\partial}^\rho \overline{\partial}^\sigma \; .
\label{Gexp2}
\end{equation}
The explicit forms for the three contractions in (\ref{Gexp0}) are,
\begin{eqnarray}
\lefteqn{4 \mathcal{D}_{\overline{\alpha\beta}}^{~~~ \mu\nu}
{\mathcal{D}'}^{\alpha\beta\rho\sigma} = \eta^{\mu\nu}
\eta^{\rho\sigma} (\nabla\!\cdot\!\nabla^\prime)^2 + \eta^{\mu\nu}
[\overline{\partial}^{\rho} \overline{\partial}^{\sigma}
{\partial^\prime}^2 \!-\! 2 \overline{\partial}^{(\rho}
\partial^{\prime\sigma)} \nabla\!\cdot\!\nabla^\prime] } \nonumber \\
& & + \overline{\eta}^{\mu(\rho} \overline{\eta}^{\sigma)\nu}
\partial^2 {\partial^\prime}^2 + \eta^{\rho\sigma}
[\overline{\partial}^{\prime\mu} \overline{\partial}^{\prime\nu}
\partial^2 \!-\! 2 \partial^{(\mu} \overline{\partial}^{\prime\nu)}
\nabla \!\cdot\! \nabla^\prime ] - 2
\overline{\partial}^{\prime(\mu} \overline{\eta}^{\nu)(\rho}
\partial^{\prime\sigma)} \partial^2 \nonumber \\
& & \hspace{2cm} - 2 \partial^{(\mu} \overline{\eta}^{\nu)(\rho}
\overline{\partial}^{\sigma)} {\partial^\prime}^2 + 2
\partial^{(\mu} \overline{\eta}^{\nu)(\rho} \partial^{\prime\sigma)}
\nabla\!\cdot\!\nabla^\prime + 2 \partial^{(\mu}
\overline{\partial}^{\prime\nu)} \overline{\partial}^{(\rho}
\partial^{\prime\sigma)} \; , \qquad \label{Gexp3} \\
\lefteqn{4 \mathcal{D}_{1 \alpha\beta}^{~~~ ~\mu\nu}
{\mathcal{D}'}^{\alpha\beta\rho\sigma} = \overline{\eta}^{\mu\nu}
\eta^{\rho\sigma} (\nabla\!\cdot\!\nabla^\prime)^2 +
\overline{\eta}^{\mu\nu} [\overline{\partial}^\rho
\overline{\partial}^\sigma {\partial^\prime}^2 - 2
\overline{\partial}^{(\rho} \partial^{\prime\sigma)} \nabla
\!\cdot\! \nabla^\prime ] } \nonumber \\
& & \hspace{0cm} + \overline{\eta}^{\mu(\rho}
\overline{\eta}^{\sigma)\nu} \nabla^2 {\partial^\prime}^2 +
\eta^{\rho\sigma} [\overline{\partial}^{\prime\mu}
\overline{\partial}^{\prime\nu} \nabla^2 - 2
\overline{\partial}^{(\mu} \overline{\partial}^{\prime\nu)} \nabla
\!\cdot\! \nabla^\prime ] - 2 \overline{\partial}^{(\mu}
\overline{\eta}^{\nu)(\rho} \overline{\partial}^{\sigma)}
{\partial^\prime}^2 \nonumber \\
& & \hspace{2cm} - 2\overline{\partial}^{\prime(\mu}
\overline{\eta}^{\nu)(\rho} \partial^{\prime\sigma)} \nabla^2 + 2
\overline{\partial}^{(\mu} \overline{\eta}^{\nu)(\rho}
\partial^{\prime\sigma)} \nabla \!\cdot\! \nabla^\prime + 2
\overline{\partial}^{(\mu} \overline{\partial}^{\prime\nu)}
\overline{\partial}^{(\rho}
\partial^{\prime\sigma)} \; , \qquad \label{Gexp4} \\
\lefteqn{4 \mathcal{D}_{\alpha\beta}^{~~ \mu\nu} {\mathcal{D}'}_1^{~
\alpha\beta\rho\sigma} = \eta^{\mu\nu} \overline{\eta}^{\rho\sigma}
(\nabla \!\cdot\! \nabla^\prime)^2 + \eta^{\mu\nu}
[\overline{\partial}^\rho \overline{\partial}^\sigma
{\nabla^\prime}^2 - 2 \overline{\partial}^{(\rho}
\overline{\partial}^{\prime\sigma)} \nabla \!\cdot\! \nabla^\prime]
} \nonumber \\
& & \hspace{0cm} + \overline{\eta}^{\mu(\rho}
\overline{\eta}^{\sigma)\nu} \partial^2{\nabla^\prime}^2 +
\overline{\eta}^{\rho\sigma} [\overline{\partial}^{\prime\mu}
\overline{\partial}^{\prime\nu} \partial^2 - 2 \partial^{(\mu}
\overline{\partial}^{\prime\nu)} \nabla \!\cdot\! \nabla^\prime] - 2
\partial^{\prime(\mu} \overline{\eta}^{\nu)(\rho}
\overline{\partial}^{\prime\sigma)} \partial^2 \nonumber \\
& & \hspace{2cm} - 2 \partial^{(\mu} \overline{\eta}^{\nu)(\rho}
\overline{\partial}^{\sigma)} {\nabla^\prime}^2 + 2
\partial^{(\mu} \overline{\eta}^{\nu)(\rho}
\overline{\partial}^{\prime\sigma)} \nabla \!\cdot\! \nabla^\prime +
2 \partial^{(\mu} \overline{\partial}^{\prime\nu)}
\overline{\partial}^{(\rho} \overline{\partial}^{\prime\sigma)} \; .
\qquad \label{Gexp5}
\end{eqnarray}

\end{document}